\documentclass[a4paper,11pt]{article}

\usepackage{graphicx,array}
\usepackage{color}
\usepackage{latexsym}
\usepackage{amsthm}
\usepackage{amsmath}
\usepackage{amssymb}
\usepackage{empheq}
\usepackage{comment}
\usepackage{dsfont}
\usepackage{epsfig}
\usepackage{slashed}
\usepackage{bbold}
\usepackage{psfrag}
\usepackage[svgnames]{xcolor}
\PassOptionsToPackage{caption=false}{subfig}
\usepackage{subcaption}
\usepackage{xfrac}
\usepackage{multirow}
\usepackage{booktabs}
\usepackage{cite}
\usepackage[normalem]{ulem}
\usepackage{jheppub} 

\newcommand{\be}{\begin{equation}}
\newcommand{\ee}{\end{equation}}
\newcommand{\bea}{\begin{eqnarray}}
\newcommand{\eea}{\end{eqnarray}}
\newcommand{\bei}{\begin{itemize}}
\newcommand{\eei}{\end{itemize}}

\newcommand{\TeV}{{\, \rm TeV}}
\newcommand{\GeV}{{\, \rm GeV}}
\newcommand{\MeV}{{\, \rm MeV}}
\newcommand{\keV}{{\, \rm keV}}

\newcommand{\fid}{{\cal F}}

\preprint{P3H-26-060, TTP26-032 }
\title{Flavor-violating dark matter at MEG-II and Mu3e}

\author[a,b]{Daniele Barducci}
\author[a,b]{Giulio Marino}
\author[a,b]{Paolo Panci}
\author[c]{Robert Ziegler}
\affiliation[a]{Dipartimento di Fisica E. Fermi, Universit\`a di Pisa, Largo B. Pontecorvo 3, I-56127 Pisa, Italy}
\affiliation[b]{INFN, Sezione di Pisa, Largo Bruno Pontecorvo 3, I-56127 Pisa, Italy}

\affiliation[c]{Institute for Theoretical Particle Physics (TTP), Karlsruhe Institute of Technology (KIT), \\
Wolfgang-Gaede-Str. 1, D-76131 Karlsruhe, Germany}
\emailAdd{daniele.barducci@pi.infn.it}
\emailAdd{giulio.marino@phd.unipi.it}
\emailAdd{paolo.panci@unipi.it}
\emailAdd{robert.ziegler@kit.edu}

\abstract{
We investigate the potential of high-intensity muon experiments such as
MEG~II and Mu3e to uncover dark matter (DM) through the lepton-flavor-violating decay $\mu^+\to e^+\chi\bar\chi$. We describe the underlying interactions in terms of dimension-six four-fermion operators and systematically explore all allowed Lorentz structures. We show that precision measurements of the Michel spectrum can probe new-physics scales of ${\cal O}({\rm TeV})$ for DM masses above approximately $1~{\rm MeV}$. For lighter DM, whose signal is confined close to the Michel endpoint, the radiative decay
$\mu^+\to e^+\chi\bar\chi\gamma$ at MEG~II opens a complementary window, with sensitivity comparable to and, in some regions, stronger than that of the non-radiative channel. Remarkably, for reheating temperatures below the tens-of-MeV scale but above the BBN bound, MEG~II and Mu3e can probe regions in which freeze-in through the very same LFV interactions accounts for the observed DM abundance.
}

\begin{document} 
\maketitle
\newpage
\section{Introduction}\label{sec:intro}
Despite the remarkable success of the Standard Model (SM), the origin of its flavor structure remains unexplained. The pattern of fermion masses and mixing suggests the presence of an underlying flavor symmetry hinting at certain dynamics beyond the SM (BSM). In this context, searches for charged lepton flavor violation (LFV) constitutes one of the most sensitive probes of BSM physics, as lepton family number is an accidental symmetry of the SM  broken only by  non-vanishing neutrino masses. The smallness of neutrino masses however implies that charged LFV processes are well beyond the reach of any foreseeable experiments~\cite{Petcov:1976ff}. Any positive signal in charged LFV searches would thus provide an unambiguous indication of New Physics (NP).

For this reason, high-intensity muon factories are currently pushing the exploration of charged LFV to unprecedented precision. In particular, the MEG-II~\cite{MEGII:2023ltw,MEGII:2025gzr} and Mu3e~\cite{Hesketh:2022wgw,Perrevoort:2018okj,Perrevoort:2018ttp,Jho:2022snj} experiments are primarily designed to search for the decay modes $\mu^+ \to e^+ \gamma$ and $\mu^+ \to e^+ e^+ e^-$, with the goal of probing NP scales that can lie beyond the energy directly accessible in collider experiments. Such processes arise naturally in a wide class of scenarios, including supersymmetric models, theories with extended gauge sectors, extra dimensional models or models with mechanisms for generating neutrino masses~\cite{Borzumati:1986qx,Ilakovac:1994kj,Hisano:1995nq,Hisano:1995cp,Kuno:1999jp,Cirigliano:2004mv,Chang:2005ag,Calibbi:2017uvl}.

At the same time the extremely large statistics and excellent detector resolution that can be achieved at modern muon facilities make these experiments also powerful laboratories for searches of light and weakly coupled particles associated with hidden or dark sectors produced in rare LFV muon decays. This is the case of axion-like particles~\cite{Bjorkeroth:2018dzu,Calibbi:2020jvd,Bauer:2021mvw,Banerjee:2022nbr, Panci:2022wlc}, sterile neutrinos~\cite{Knapen:2024fvh} or stable states playing the role of Dark Matter (DM)~\cite{Jahedi:2025hnu,Roig:2026wvo}. 

\medskip
In this work, we investigate a Dirac fermion DM candidate, $\chi$, interacting
with the SM through LFV dimension-six four-fermion operators of the schematic
form $(\bar e\Gamma_\ell\mu)(\bar\chi\Gamma_\chi\chi)$. Within an effective
field theory framework, we classify all possible Lorentz structures built
from SM-lepton and DM bilinears, and assess their prospects at MEG~II and
Mu3e for DM masses below the kinematic threshold of muon decay. In this mass
range, conventional direct-detection searches have limited sensitivity,
since these LFV operators do not mediate elastic DM scattering off electrons
or nuclei at tree level. The corresponding inelastic process
$\chi e\to\chi\mu$ is kinematically inaccessible to non-relativistic Galactic
DM, while elastic scattering can arise only through loop-induced and
model-dependent interactions. At the same time, the annihilation channel
$\chi\bar\chi\to e\mu$ relevant for indirect detection is kinematically
forbidden. Rare muon-decay searches at high-intensity facilities therefore
provide a particularly direct and sensitive probe of lepton-flavor-violating
DM interactions.

We consider both the cases of $\mu^+ \to e^+ \chi \bar{\chi}$ and  $\mu^+ \to e^+ \chi \bar{\chi}\gamma$ decays and perform a shape analysis against the SM predictions for these processes, showing that 
the radiative one offers an additional handle for testing light DM masses, for which the BSM signal lies close to the Michel's spectrum endpoint.
We find that precision measurements at muon factories such as  MEG-II and Mu3e can test effective scales of ${\cal O}({\rm TeV})$, which is comparable to astrophysical constraints from supernova cooling.  
We further discuss the cosmological implications of this scenario, showing that for  reheating temperatures below $\sim10\,\mathrm{MeV}$ but above the big bang nucleosynthesis (BBN) bound, the parameter space accessible to MEG-II and Mu3e overlaps with regions in which the observed DM relic abundance can be generated through freeze-in  via the very same  effective LFV interactions.

\medskip
The rest of the paper is organised as follows. In Sec.~\ref{sec:thframe} we introduce the EFT for charged LFV interactions and compute the differential $\mu$ decay rates for the various Lorentz structures of the four-fermion operators. In Sec.~\ref{sec:inv_decay} we present our analysis strategy and study the $\mu^+ \to e^+\chi\bar \chi$ decay, while Sec.~\ref{sec:mutoegamma} is dedicated to the analysis of the radiative $\mu^+ \to e^+\chi\bar \chi\gamma$ process. In Sec.~\ref{sec:complementary} we discuss other complementary constraints while in Sec.~\ref{sec: Thermal production} we analyse the cosmological properties of the DM candidate $\chi$. We conclude in Sec.~\ref{sec:conclusions}. We also add four appendices with some technical details and additional results not presented in the main text.

\section{Theoretical framework}\label{sec:thframe} 
We extend the low-energy SM by introducing a dark Dirac fermion, denoted by $\chi$, which is a singlet under the SM group. We assume that $\chi$ is stabilized by a discrete $\mathbb{Z}_2$ symmetry, which forbids operators containing an odd number of $\chi$ fields. The EFT adopted in this work is defined at the relevant energy scale of LFV experiments, where the typical momenta involved in stopped muon decays are of order the muon mass. At these energies, the relevant SM degrees of freedom for the process shown in Fig.~\ref{fig:feyndiag} are the electron and the muon, to which we add the DM candidate $\chi$. We therefore describe possible short distance SM--DM interactions in terms of local four-fermion operators. In this framework, the LFV decay
\be\label{eq:LFV_decay}
\mu^+ \to e^+ \chi \bar \chi \ ,
\ee
can be induced by $d=6$ operators involving one muon, one electron, and two DM fields. The most general four-fermion interactions can be organized according to the Lorentz structure of the fermion bilinears. 
{The corresponding effective interaction Lagrangian can be written as

\begin{equation}\label{eq:lag}
\mathcal{L}_{\rm LFV}  = \frac{1}{\Lambda^2} \sum_{X,Y}
\big( \bar e  c^\ell_{X} \Gamma_{X}  \mu\big)
\big(\bar \chi  c^\chi_{Y} \Gamma_{Y} \chi\big) + \text{h.c.} \, , 
\end{equation}
where $\Lambda$ is the cutoff scale of the EFT and $X,Y$ run over the combinations of fermion bilinears that yield Lorentz-invariant four-fermion operators, namely $S,P,V,A,T,T^\prime$ with 
$\Gamma_{X} = \{ \mathbb{1}, \gamma_5, \gamma_\mu, \gamma_\mu \gamma_5, \sigma_{\mu\nu}, \sigma_{\mu\nu} \gamma_5\}$. These operators describe scalar, pseudoscalar, vector, axial-vector, tensor, and pseudotensor LFV and DM currents, respectively\footnote{The tensor interaction contains only two independent structures. Since, in four dimensions, the pseudotensor current is related to the tensor one through the identity involving $\sigma^{\mu\nu}\gamma^5$, one of the two can be eliminated without loss of generality. In our convention we choose to set the DM pseudotensor coefficient to zero. Further details are given in App.~\ref{eq:tensor_coeff}.}.  Here $c_X^{\ell}$ and $c_X^{\chi}$ are in general complex Wilson coefficients related to the LFV fermion and DM bilinears, respectively. In this work, we focus on scenarios in which the DM and leptonic sectors share the same bilinear structure, \textit{i.e.} $X = Y$, since for small DM masses LFV probes are insensitive to the invisible sector. Hence, we define the effective scale $\Lambda^{\mu e}_{X}\equiv\Lambda/\sqrt{c_{X}^\ell c_{X}^\chi}$. 

\medskip
Although Eq.~\eqref{eq:lag} provides the most general low-energy parametrization relevant for LFV experiments, not all Lorentz structures are equally minimal from the point of view of a weakly coupled, EW gauge invariant ultraviolet (UV) completion. For example vector interactions can arise rather directly from the exchange of a heavy neutral mediator coupled both to a LFV current and to the DM current. Scalar interactions are already less minimal, since EW gauge invariance often requires Higgs insertions or additional sources of chiral symmetry breaking. For instance, scalar LFV structures involving left- and right-handed charged leptons can arise schematically from operators of the form $(\bar L H \mu_R)\,\bar\chi\chi$, which are $d=7$ before EW symmetry breaking and reduce to scalar four-fermion interactions only after the Higgs acquires its vacuum expectation value. Tensor interactions are even less minimal. Unlike scalar or vector structures, they do not arise in isolation from the tree level exchange of a fundamental weakly coupled mediator with renormalizable interactions.
Rather, they should be viewed as genuinely effective structures, typically generated only after integrating out a more complicated mediator sector or through loop effects and additional insertions. Their UV origin is therefore intrinsically more model-dependent.
For this reason, while we keep the general low-energy basis of Eq.~\eqref{eq:lag} in order to describe all possible kinematic structures of the decay, the vector interaction will play a central role in the following phenomenological analysis and will be discussed in the main text. Scalar and tensor benchmark scenarios are nevertheless discussed in App.\,\ref{app:distr_tensor} in order to illustrate how results from LFV experiments depend on the Lorentz structure of the underlying interaction.

\begin{figure}[t!]
  \centering
  \includegraphics[width=0.45\textwidth]{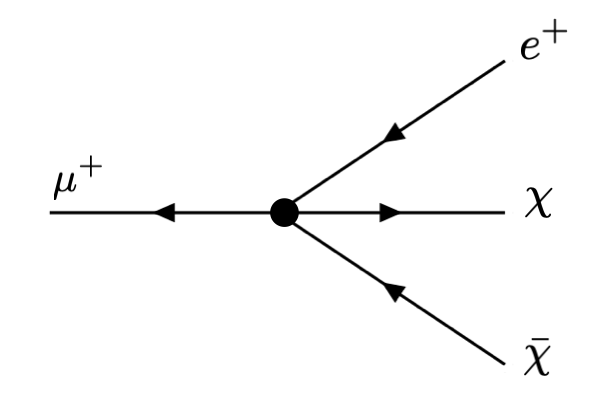}
  \caption{Feynman diagram for the LFV $\mu^+\to e^+\bar\chi\chi$ process. The black dot represents the insertion of the effective operators of Eq.~\eqref{eq:lag}.}
  \label{fig:feyndiag}
\end{figure}

From the interactions of Eq.~\eqref{eq:lag} we can compute the differential decay rates for a muon $\mu^+$ with degree of polarization $\cal P$ quantized along the muon beam  direction. Upon integrating over the unobservable DM momenta, one has in the limit of vanishing positron and DM masses the following expressions (see App.\,\ref{app:mu-decay} for the full $m_\chi$-dependence) 
\begin{align}
\label{eq:rates}
\frac{{\rm d^2}\Gamma_{\rm scalar}}{{\rm d}E_e\,{\rm d}\cos\theta_e}
&= \frac{48 E_e^2}{m_\mu^3 } \Gamma_{\rm scalar}   \left(1-2 \,\frac{E_e}{m_\mu}\right) 
\left(
1
+2{\cal P}\cos\theta_e\,
\frac{{\rm Re}\,[c_S^\ell c_P^{\ell*}]}{|c_S^\ell|^2+|c_P^\ell|^2}
\right) \,  \\
\frac{{\rm d^2}\Gamma_{\rm vector}}{{\rm d}E_e\,{\rm d}\cos\theta_e}
&= \frac{8 E_e^2}{m_\mu^3 } \Gamma_{\rm vector}    \left(3-4 \,\frac{E_e}{m_\mu}\right) 
\left( 1
+2{\cal P}\cos\theta_e\,
\frac{{\rm Re}\, [c_V^\ell c_A^{\ell*}]}{|c_V^\ell|^2+|c_A^\ell|^2} \frac{1-4 \frac{E_e}{m_\mu}}{3 -4 \frac{E_e}{m_\mu}} \right) \ ,  \\
\frac{{\rm d^2}\Gamma_{\rm tensor}}{{\rm d}E_e\,{\rm d}\cos\theta_e}
&=  \frac{16 E_e^2}{3 m_\mu^3 } \Gamma_{\rm tensor}  \left(3-2 \,\frac{E_e}{m_\mu}\right)   
\left(
1 -2{\cal P}\cos\theta_e \,
\frac{{\rm Re}[c_{T}^\ell c_{T^\prime}^{\ell*}]}{|c_{T}^\ell|^2+|c_{T^\prime}^\ell|^2} \frac{1 + 2 \frac{E_e}{m_\mu}}{3 -2 \frac{E_e}{m_\mu}} \right)\, , 
\end{align}
where $\theta_e$ is the angle between the emitted positron and the beam axis, $E_e$ with $0\leq E_e\leq m_\mu/2$
 is the positron energy  and the total rates read
\begin{equation}\label{eq:totalrates}
\begin{split}
\Gamma_{\rm scalar} & = \big(
|c_S^\chi|^2 
+|c_P^\chi|^2 
\big)
\big(
|c_S^\ell|^2+|c_P^\ell|^2 \big) \frac{m_\mu^5}{1536 \pi^3 \Lambda^4} \, ,  \\
\Gamma_{\rm vector} & = \big(
|c_V^\chi|^2 
+|c_A^\chi|^2 
\big)
\big(
|c_V^\ell|^2+|c_A^\ell|^2 \big) \frac{m_\mu^5}{384 \pi^3 \Lambda^4} \, , \\
\Gamma_{\rm tensor} & = 
|c_T^\chi|^2 
\big(
|c_T^\ell|^2+|c_{T^\prime}^\ell|^2 \big) \frac{m_\mu^5}{64 \pi^3 \Lambda^4} \, . 
\end{split}
\end{equation}
We fix the muon polarization to ${\cal P}=-0.85$ corresponding to the value measured in the MEG-II experiment~\cite{MEG:2015kvn}, which we also adopt for Mu3e. For concreteness, for the vector interaction we impose one of the following conditions: $c_V^{\ell,\chi}=0$, $c_A^{\ell,\chi}=0$, or $c_V^{\ell,\chi}=\pm c_A^{\ell,\chi}$. The last choice corresponds to a change of basis, in which the associated Wilson coefficients are defined as  
$c_R^{\ell,\chi}/\sqrt{2}\equiv c^{\ell ,\chi}_V=c^{\ell ,\chi}_A $ 
or 
$c_L^{\ell,\chi}/\sqrt{2}\equiv c^{\ell ,\chi}_V=-c^{\ell ,\chi}_A $. As a result, the corresponding  effective scales in the left/right basis read $\Lambda_{L,R}^{\mu e}\equiv \Lambda/\sqrt{c^{\ell}_{L,R}c^{\chi}_{L,R}}$. In Fig.~\ref{fig:diffrates} we show the differential decay rates in $E_e$ and $\cos\theta_e$, integrated over $\cos\theta_e$ and $E_e$ respectively, for the vector and scalar type of interactions and for different choices of DM mass, while the corresponding plots for the tensor interactions are shown in App.\,\ref{app:distr_tensor}. 

We observe that the differential distribution in positron energy is largely insensitive to Lorentz structure. This is expected, as in the massless DM limit the differential rates  are identical upon integration on the $\theta_e$ angle, as evident from Eq.~\eqref{eq:rates}, with small differences arising from $m_\chi$-corrections, see Eq.~\eqref{eq:LFVtotal_scalar}. Moreover, in the massless DM limit one also obtains a NP spectrum identical to the SM Michel decay, which is represented by the black solid line, independently on the effective scale.

\begin{figure}[h!]
  \centering
  \includegraphics[width=0.48\textwidth]{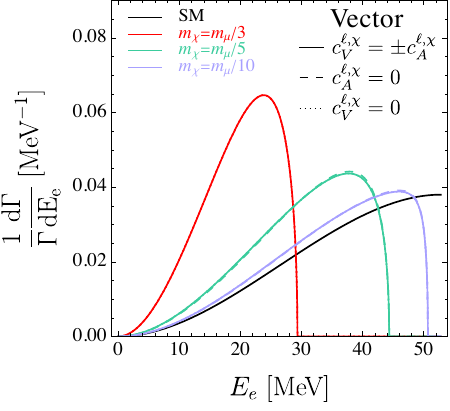}\hfill
  \includegraphics[width=0.48\textwidth]{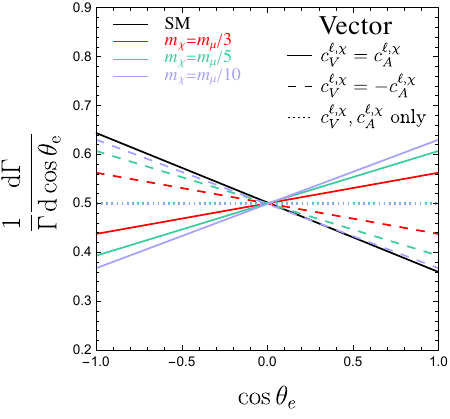}
  \caption{Differential rates as function of $E_e$ and inclusive in $\cos\theta_e$ (left panel) and  as function of $\cos\theta_e$ and inclusive in $E_e$ (right panel) for the vector case. In all the panels we fix the muon polarization as ${\cal P}=-0.85$. For the vector case we fix $c_V^{\ell,\chi}=0$, $c_A^{\ell,\chi}=0$ or  $c_V^{\ell,\chi}=\pm c_A^{\ell,\chi}$.
 The red, green and purple curves correspond to $m_\chi = m_\mu/3,\, m_\mu/5$ and  $m_\mu/10$ respectively, while the black solid line represents the SM Michel spectrum. }
  \label{fig:diffrates}
\end{figure}

Moving to the $\cos\theta_e$ distributions provides a handle to distinguish between the different chiralities. We see that for the pure left-handed scenario, \textit{i.e.} $c_V^{\ell\chi} = - c_A^{\ell \chi}$, the NP spectrum approaches the Michel decay when decreasing the DM mass, making the vector type interaction indistinguishable from the SM in both energy and angular distribution in the $m_\chi\to0$ limit. On the other hand, the angular distribution in the pure right-handed case, \textit{i.e.} $c_V^{\ell\chi} = + c_A^{\ell \chi}$, has an opposite slope with respect to the SM one. This offers a clear handle to distinguish the signal with respect to the SM background also in the case of small DM mass. Finally, in the pure vector or axial case, \textit{i.e.}, when $c_A^\ell$ or $c_V^\ell$ vanishes, respectively, the decay is isotropic and the angular distribution remains flat for all DM masses. Even in this case, however, the signal exhibits a shape clearly distinct from the SM background.

\section{Rare LFV decay at MEG-II and Mu3e}\label{sec:inv_decay}

Having illustrated the general structure of the rare LFV muon decay induced by the operators of Eq.~\eqref{eq:lag}, in this Section we assess the sensitivity on the NP effective scales $\Lambda^{\mu e}_{X}$ that can be attained by a precise measurement of the SM Michel spectrum performed at MEG-II and Mu3e. For clarity of presentation, in the main text we focus only on vector-type interactions and consider three benchmark scenarios: $X=L$, corresponding to a pure left-handed (LH) coupling with $c_V^{\ell,\chi}=-c_A^{\ell,\chi}$; $X=R$, corresponding to a pure right-handed (RH) coupling with $c_V^{\ell,\chi}=c_A^{\ell,\chi}$; and $X=V(A)$, corresponding to the pure vector (axial) case, in which $c_A^{\ell,\chi}$ ($c_V^{\ell,\chi}$) vanishes. Since the pure vector and pure axial cases differ only by subleading corrections of order $m_\chi$, they lead to essentially identical phenomenology and are therefore treated as a single benchmark. Results for the scalar and tensor operators are presented in App.~\ref{app:distr_tensor}. Throughout the analysis, we restrict ourselves to real couplings.

We first describe the general analysis strategy and statistical procedure adopted, and then discuss the reach of the two experiments in turn.

\subsection{Analysis strategy and statistical procedure}\label{sec:strategy}

For our analysis we choose to only exploit the shape difference between the SM and the NP spectrum within the experimental fiducial region $\fid$, defined as
\be
\fid = \left\{ (E_e, \cos\theta_e)\;\middle|\;
E_{e,\min} < E_e < E_{e,\max},\;
\cos\theta_{e,\min} < \cos\theta_e < \cos\theta_{e,\max}
\right\} \ ,
\ee
 and not the overall event rate normalisation\footnote{The MEG-II and Mu3e experiment also select events based on positron azimuthal angle $\varphi_e$. This however only reflects on an overall reduction of the total measured rate, which we take into account with a flat rescaling factor of the integrated dataset}.  This strategy avoids situations in which a bound can be extracted even if the NP signal lies entirely outside $\fid$, {\emph{e.g.}} in the case of large DM mass for which the positron energy spectrum might lie outside $\fid$, by the measurement of a reduction of the SM rate. In other words we fix the total number of decayed muons measured by the experiment $N_{\mu^+}$ to the SM expectation, assuming that this rate is fixed through a short calibration phase of the experiment.

For our analysis we then proceed as follows. Within $\fid$ we define a total probability distribution function (PDF) 
\be\label{eq:pdf}
f_{\rm TOT}(E_e,\cos\theta_e,m_\chi,\alpha)  = \alpha f_{\rm NP}(E_e,\cos\theta_e,m_\chi) + (1-\alpha) f_{\rm SM}(E_e,\cos\theta_e) \ , 
\ee 
as a weighted sum of the NP and SM contributions, where
\be
\alpha = \frac{N^{\rm NP}_{\fid}}{N_{\fid}^{\rm TOT}} \ ,  \quad 0 \le \alpha \le 1
\ee
is the ratio of the NP events over the total observed decayed muon within $\fid$. The NP and SM PDFs entering Eq.~\eqref{eq:pdf} can be built normalising the rates of Eq.~\eqref{eq:rates} within $\fid$.

To model the detector response we then convolute the theoretical rates with the efficiency for positron reconstruction ${\cal A}_e$, which we assume to depend only on the positron energy and that we directly extract from the experimental setups and with a detector response function  ${\cal S}$ modeled, following~\cite{Banerjee:2022nbr}, as a gaussian distribution  with zero mean and standard deviation $\sigma_S$
\begin{equation}
    \mathcal{S}_e(E_e,E_e')
    =
    \frac{1}{\sigma_{\mathcal{S}_e}(E_e^\prime)\sqrt{2\pi}}
    \exp\!\left[
        -\frac{1}{2}
        \left(
            \frac{E_e-E_e^\prime}{\sigma_{\mathcal{S}_e}(E_e^\prime)}
        \right)^2
    \right] \ ,
\end{equation}
where $E_e$ represents the positron energy. The convoluted differential rates are then computed as
\be
    {\cal J}_i(E_e,\cos\theta_e,m_\chi)
    =
    \int dE_e^\prime\,
    \left(
        \frac{{\rm d}\Gamma_i}{{\rm d}E_e^\prime\,{\rm d}\cos\theta_e}
        \times \mathcal{A}_e(E_e^\prime)
        \times \mathcal{S}_e(E_e,E_e^\prime)
    \right) \ ,
\label{eq:spectra}
\ee
where $i$ indicates either of the SM or NP contributions. The PDFs entering in Eq.~\eqref{eq:pdf} are then
\be\label{eq:pdf_SM_NP}
f_i(E_e,\cos\theta_e,m_\chi) = \frac{{\cal J}_i(E_e,\cos\theta_e,m_\chi)}{\int_{\cal F}dE_e d\cos\theta_e {\cal J}(E_e,\cos\theta_e,m_\chi)} \ ,
\ee
thus ensuring that $f_{\rm TOT}(E_e,\cos\theta_e,m_\chi,\alpha)$ is properly normalized to unity. We then divide the fiducial region $\fid$ into $ N_{\rm bin} = N_{{\rm bin},\,E_e} \times N_{{\rm bin},\,\cos\theta_e}$ equally spaced bins  and compute, by integrating Eq.~\eqref{eq:pdf_SM_NP} and Eq.~\eqref{eq:pdf} and multiplying them by the total number of observed decayed muons $N_{\mu^+}$, the total and SM expected rates $N_{{\rm TOT},i}$ and $N_{{\rm SM},i}$, where $i=1,\dots,N^{\rm bin}$. We then build a $\chi^2$ distribution
\be\label{eq:chi2}
\chi^2(\alpha)  = \sum_{i=1}^{N_{\rm bin} } \frac{{\big[N_{{\rm TOT},i}-N_{{\rm SM},i}\big]^2}}{
N_{{\rm TOT},i}+ (\varepsilon_{\rm syst} N_{{\rm TOT},i})^2} \ ,
\ee
where with $\varepsilon_{\rm syst}$ we parametrize the experimental systematic uncertainty and $\sum_i N_{{\rm SM},i}=N_{\mu^+}$. In accord with our assumptions of normalizing the total rate to the expected SM one, the $\chi^2$ vanishes in the case of a single bin. It also vanishes in the case that the SM and NP signals have the same shape within the chosen binning, since the integral of Eq.~\eqref{eq:pdf_SM_NP} and Eq.~\eqref{eq:pdf} within each bin coincide. Clearly this happens for massless DM in the pure LH scenario, where the SM and NP spectra coincide in both positron energy and polar angle, see the discussion in Sec.~\ref{sec:thframe}.

Once an upper limit on $\alpha$ is obtained, it can be translated into a
bound on the model parameters by explicitly expressing  $\alpha$ as
\be\label{eq:alpha}
\alpha = \frac{
{\rm BR}_{\rm NP} {\cal R}_{{\rm NP}}
}
{
{\rm BR}_{\rm SM} {\cal R}_{{\rm SM}}+{\rm BR}_{\rm NP} {\cal R}_{{\rm NP}}
} \ ,
\ee
where  ${\rm BR}_{\rm SM}$ and ${\rm BR}_{\rm NP}$ are the $\mu^+$ branching ratio (BR) into the NP and SM final state and we assume ${\rm BR}_{\rm SM}+{\rm BR}_{\rm NP}=1$, while ${\cal R}_{{\rm NP},\,{\rm SM}}$ represent the fraction of experimentally detected positrons over the total number of positron for both the SM and NP cases
\be
\begin{split}
 {\cal R}_{\rm SM}  & =\frac{1}{\Gamma_{\rm SM}^{\rm TOT}}\int_\fid {\rm d}E_e \, d \cos\theta_e \,  {\cal J}_{\rm SM}(E_e,\cos\theta_e,m_\chi) \ , \\
{\cal R}_{\rm NP} & = \frac{1}{\Gamma_{\rm NP}^{\rm TOT}}\int_{\fid} {\rm d}E_e \, d \cos\theta_e \, {\cal J}_{\rm NP}(E_e,\cos\theta_e,m_\chi)  \ .
\end{split}
\ee 
With these definitions, and in the limit of an ideal detector covering the whole solid angle, the parameter $\alpha$ of Eq.~\eqref{eq:alpha} represents the $\mu^+$ BR into the NP final state.

\subsection{Analysis at MEG-II}\label{sec:MEG-II_inv}

MEG-II~\cite{MEGII:2018kmf} is an experiment located at the Paul Scherrer Institute (PSI) whose primary goal is to search for the LFV decay $\mu^+\to e^+\gamma$. The SM prediction for this BR is at the level of ${\cal O}(10^{-50})$ and its observation will then be a clear sign of NP. The current limit at 95\% confidence level (CL) for this rate is $1.5\times 10^{-13}$~\cite{MEGII:2025gzr}, which is expected to improve down to $\sim 6\times 10^{-14}$ with future data~\cite{MEGII:2018kmf}. MEG-II employs a high intensity $\mu^+$ beam, stopped in a thin target located at the center of a magnetic spectrometer, with a high resolution liquid xenon scintillation detector and a drift chamber measuring the final photon and positron respectively.
The MEG-II fiducial region $\fid$ that we adopt for our analysis is defined by $40\,{\rm MeV} < E_e < {m_\mu}/{2}$, $|\cos \theta_e|<0.35$ and ${2\pi}/{3}<\phi_e<{4\pi}/{3}$~\cite{MEGII:2023fog}.  We parametrize the efficiency ${\cal A}_e$ for positron reconstruction be extracting the experimental distributions reported in Ref.~\cite{MEGII:2023fog} while, following Ref.~\cite{Banerjee:2022nbr},  we choose a constant value for the uncertainty on the detector response function $\sigma_{\mathcal{S}_e} = 0.1~{\rm MeV}$. The muon polarization fraction is instead fixed to the measured value ${\cal P}=-0.85$~\cite{MEG:2015kvn} while the  muon stopping rate is $R_{\mu^+}=3\times10^7\,{\rm s}^{-1}$~\cite{MEGII:2023ltw}, corresponding to a dataset of $\simeq10^{14}$ integrated over one year of data taking. 

The goal of our analysis is to constrain the interactions of Eq.~\eqref{eq:lag} by using the SM Michel spectrum distribution. The MEG-II experiment, however, by design only records event where a $e^+\gamma$ pair are measured in coincidence. This means that, in principle, a pure Michel decay is not directly observed in the experiment. Nevertheless, as will be discussed in more detail in Sec.~\ref{sec:mutoegamma} when we will analyse the radiative $\mu^+ \to e^+\chi\bar \chi \gamma$ process, due to the 
high intensity of the $\mu^+$ beam the dominant SM background at MEG-II arises from random coincidence (RC) events, in which a Michel positron is paired up with an uncorrelated photon originating either from a radiative muon decay (RMD) $\mu^+\to e^+ \nu_e \bar \nu_\mu \gamma$
or from positron annihilating into a pair of photons in the detector material, where one of the two photons is lost. This means that the positron spectrum of the $e^+\gamma$ events measured by MEG-II can be considered to be the SM Michel one, which we will therefore use in our shape analysis. The probability for having this RC pairing between positron and photons, both entering the MEG-II fiducial region, imply that the total number of available muon for our analysis is smaller than the number of stopped muons. We will treat the normalization of the Michel spectrum as a free parameter, which we vary in the range $N_{\mu^+}\simeq10^{5}-10^7$. This is consistent with a direct estimate that we will discuss in more detail in Sec.~\ref{sec:mutoegamma} when thoroughly discussing the radiative $\mu^+\to e^+\chi \bar \chi \gamma$ process.

We now apply the  strategy discussed in Sec.~\ref{sec:strategy}. We choose $N_{{\rm bin},E_e}=N_{{\rm bin},\cos\theta_e}=10$ equally spaced bins and show the
limit on the parameter $\alpha$ defined in  Eq.~\eqref{eq:alpha} in function of the DM mass in Fig.~\ref{fig:alphabounds} for the case of vector bilinear, with different assumptions on the interaction coupling structure. We compare the results that we obtain by only binning in the positron energy $E_{e}$, positron polar angle $\cos_{\theta_e}$ or both, and show the limits for $N_{\mu^+}=10^6$, both under the assumptions of either vanishing systematic uncertainties or by fixing $\varepsilon_{\rm syst}=10^{-3}$, roughly corresponding to the MEG-II uncertainty on positron reconstruction~\cite{MEGII:2023ltw,MEGII:2025gzr}. We see that the limits do not change appreciably by varying $\varepsilon_{\rm syst}$, consistently with a rough estimate of the average statistical uncertainty in each bin of order of $(N_{\mu^+}^i)^{-1/2}\simeq 10^{-2}$ with $N_{\mu^+}^i\simeq N_{\mu^+}/N_{\rm bin}=10^4$, meaning that the uncertainties of the proposed analysis are statistically dominated.

In the top panels of Fig.~\ref{fig:alphabounds} we show the 95\% CL lower bounds on $\alpha$ as a function of the DM mass for the three benchmark scenarios discussed above: pure RH $c_V^{\ell,\chi}=c_A^{\ell,\chi}$ (blue lines), pure LH $c_V^{\ell,\chi}=-c_A^{\ell,\chi}$ (orange lines) and pure vector or axial scenarios (purple lines). The two panels correspond to two different binning choices: binning only in $\cos\theta_e$ (left panel), and binning only in positron energy (right panel). Solid and dashed lines correspond to the cases with and without systematic uncertainties, respectively. The three benchmark choices of couplings lead to identical constraints when binning only in the positron energy, as shown in the right panel. This occurs because, as shown in Eq.~\eqref{eq:rates}, the only term sensitive to the relative sign of $c_V$ and $c_A$ is the one proportional to $\cos\theta_e$. In contrast, in the pure vector or axial scenarios the coupling scaling is identical, and the corresponding distributions are flat in $\cos\theta_e$. When looking at the differential rate for the vector interaction in Fig.~\ref{fig:diffrates} we highlighted that for vanishing DM mass the NP positron energy spectrum becomes degenerate with the SM Michel one. Still from Fig.~\ref{fig:diffrates}, we also see that the smaller $m_\chi$, the closer the NP distortion lies to the Michel endpoint $E_e = {m_\mu}/{2}$. However this region of the positron energy spectrum is usually assumed to be signal free by MEG-II, and used for calibration purposes. For this region in Fig.~\ref{fig:alphabounds} we shade in gray the region of DM mass where the NP distortion lies within $100\keV$ of the Michel endpoint, corresponding to a positron energy resolution of $2\times 10^{-3}$~\cite{Tassielli:2020wap}, where our bound cannot be trusted, which roughly correspond to $m_\chi \simeq 1\,$MeV (a strategy to constraint lighter DM masses via RMD will be discussed in Sec.~\ref{sec:mutoegamma}).
The limits on $\alpha$ can then be mapped onto a bound on the  NP scale $\Lambda$ by using Eq.~\eqref{eq:alpha}. The results are shown in the left panel of Fig.~\ref{fig:Bound_inv_search}. Here, we present the results by fixing the systematic uncertainties at the permille level and varying the number of muons from $10^5$ to $10^7$ (indicated by the shaded bands, with the central value shown as a solid line). The constraints on $\Lambda^{\mu e}_X$ are displayed for $X=R$ in blue, $X=L$ in orange and pure vector or axial scenario in purple. The results show that for the pure RH combination the bound is flat and of $\mathcal{O}(\mathrm{TeV})$, while for the LH combination the bound becomes weaker at lower we, since in the limit $m_\chi \to 0$ the NP signal is completely degenerate with the SM one. For the pure vector or axial scenario, the bound remains flat, although weaker than in the RH scenario. As can be seen from the differential rates in Fig.~\ref{fig:diffrates}, the $X=V$ or $A$ cases yield a flat distribution in $\cos\theta_e$, which provides less constraining power than the $X=R$ scenario, whose slope is opposite to the Michel behavior. 

As anticipated by the results in Fig.~\ref{fig:alphabounds}, the bounds shown in Fig.~\ref{fig:Bound_inv_search} indicate that for small values of $m_\chi$ the sensitivity is mainly driven by the angular information.
In this regime, the $X=R$ scenario provides a stronger discrimination power than the $X=V,A$ cases, leading to a constraint that is approximately $20\%$ more stringent. Since the spectral and angular distributions become essentially independent of $m_\chi$ in the small-mass limit, the resulting bounds approach a constant value and no distinctive mass-dependent features are expected. The sensitivity reaches its maximum for $m_\chi\sim\mathcal{O}(10~\mathrm{MeV})$, where both the angular and positron-energy distributions contribute significantly to the discrimination between the NP signal and the SM background. In this region, the bounds become progressively less dependent on the coupling scenario, as the information provided by the positron-energy distribution is insensitive to the relative vector and axial structure of the interaction. At larger DM masses, the three scenarios therefore yield comparable sensitivities, up to the kinematic endpoint where the available phase space closes.

\begin{figure}[t!]
    \centering
    \includegraphics[width=1\linewidth]{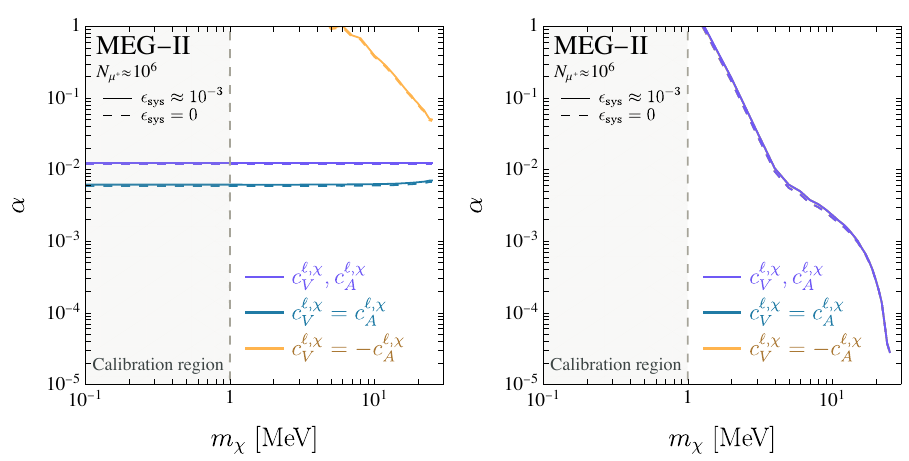}\\
    \includegraphics[width=1\linewidth]{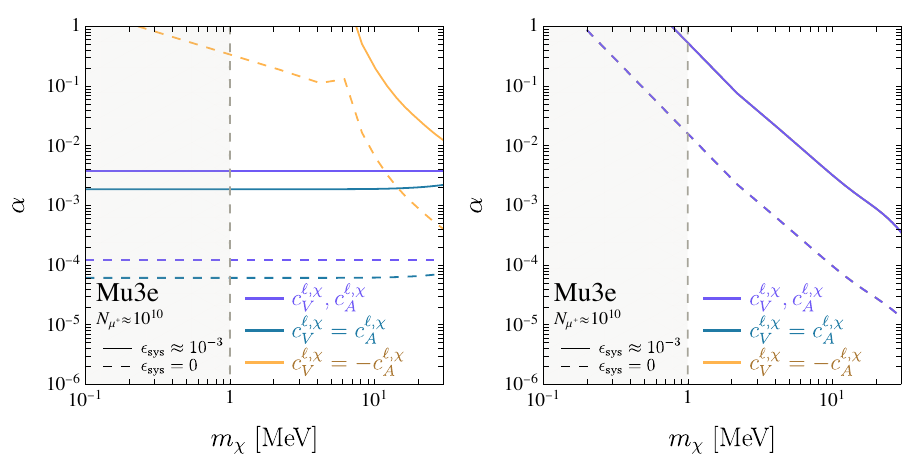}\\
    \caption{
   \textbf{Top panels:} 95\% CL constraints on $\alpha$ as a function of the DM mass for $c_V^{\ell,\chi}=c_A^{\ell,\chi}$ (blue lines), $c_V^{\ell,\chi}=-c_A^{\ell,\chi}$ (orange lines) and pure vector or axial couplings (purple lines), shown for two different binning choices: (left panel) binning only in $\cos\theta_e$ (10 bins), and (right panel) binning only in positron energy (10 bins). Solid lines correspond to the inclusion of systematic uncertainties at the permille level, while dashed lines represent the case with no systematic uncertainties.
   \textbf{Bottom panels:} 95\% C.L constraints of $\alpha$ as a function of the DM mass for Mu3e. The description is the same as in the top panels, with the only difference being the number of muons, which for Mu3e is  $N_{\mu^+} = 10^{10}$. For both MEG-II and Mu3e, in the second column corresponding to the $\alpha$ binning in positron energy, the solid blue, orange, and purple curves are completely overlapped.}
    \label{fig:alphabounds}
\end{figure}

\begin{figure}[t!]
    \centering
    \includegraphics[width=0.49\linewidth]{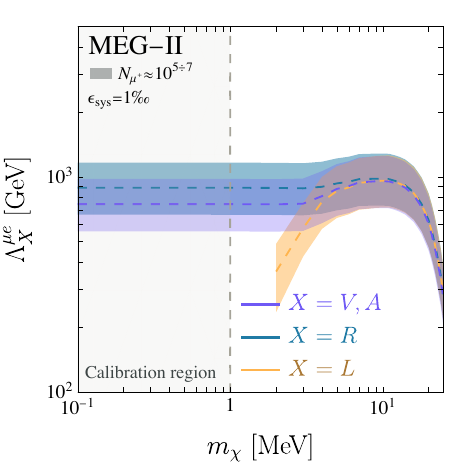}
    \hfill
    \includegraphics[width=0.49\linewidth]{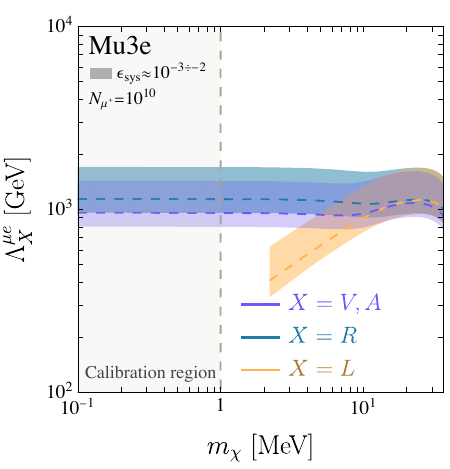}
    \caption{95\% CL constraints from invisible searches at MEG-II (left panel) and Mu3e (right panel) on the new physics effective scale $\Lambda^{\mu e}_X$ for three benchmark scenarios: pure right-handed $c_V^{\ell,\chi}=c_A^{\ell,\chi}$ (blue lines), pure left-handed $c_V^{\ell,\chi}=-c_A^{\ell,\chi}$ (orange lines) and pure vector and axial (light purple lines). For MEG-II, we fix the systematic uncertainty at the permille level and show the constraints for a number of muons ranging from $10^5$ to $10^7$. For Mu3e, we fix the number of muons to $10^{10}$ and present the results by varying the systematic uncertainty between $10^{-3}$ and $10^{-2}$. In both experimental setups, the results are obtained by binning in both positron energy and $\cos\theta_e$.  
    }
    \label{fig:Bound_inv_search}
\end {figure}

\subsection{Analysis at Mu3e}
\label{subsec:mu3eanalysis}

Mu3e~\cite{Mu3e:2020gyw} is a proposed experiment at PSI 
with the goal of measuring the $\mu^+ \to e^+e^+ e^-$ LFV muon decay, which similar to the case of $\mu^+ \to e^+\gamma$ has an almost vanishing  branching ratio of ${\cal O}(10^{-50})$ in the SM, making it an exquisite clean probe of BSM physics. The current limit is ${\rm BR} (\mu^+ \to e^+e^+ e^-) \le 1.0\times 10^{-12}$ from the SINDRUM experiment~\cite{SINDRUM:1987nra}, which will be improved by roughly four orders of magnitude with the full Mu3e dataset~\cite{Perrevoort:2023qhn}. 

Mu3e employs a high-intensity $\mu^+$ beam stopped in a thin double hollow cone target. High precision tracking and timing detectors allow for a precise reconstruction of the final state electrons and positrons~\cite{Mu3e:2020gyw,Hesketh:2022wgw}. The Mu3e fiducial region that we adopt for our analysis is defined by $15\,{\rm MeV} < E_e < {m_\mu}/{2}$, $|\cos \theta_e|<0.8$ and no restriction on $\phi_e$ is taken. 
Following Ref.~\cite{Banerjee:2022nbr} we  assume an energy-dependent uncertainty for the detector response function $\sigma_{{\cal S}_e}=0.05 E_e$  and take the positron reconstruction efficiency ${\cal A}_e$ from the parametrization of Ref.~\cite{Banerjee:2022nbr}, based on the detector performance reported in Ref.~\cite{Mu3e:2020gyw}. With respect to MEG-II, the Mu3e experiment has a larger fiducial region $\fid$ at the price a smaller energy resolution.  With a rate of $R_{\mu^+}\simeq 10^8\,{\rm s}^{-1}$ stopped $\mu^+$, Mu3e aims at collecting a dataset of ${\cal O}(10^{16})$ muons by the end of its operations~\cite{Perrevoort:2018ttp}. As for the MEG-II analysis described in Sec.~\ref{sec:MEG-II_inv}, also here our goal is to constraint the interactions of Eq.~\eqref{eq:lag} by a shape analysis of the SM Michel spectrum.
Background processes in Mu3e mainly arise from radiative $\mu^+$ decay with internal conversion $\mu^+ \to e^+ e^+ e^- \bar \nu_\mu \nu_e$ and accidental backgrounds from the combination of one or more Michel decays with positron arising from Bhabha scattering or radiative $\mu^+$ decay followed by photon conversion in the detector material~\cite{Hesketh:2022wgw}. Differently from MEG-II the recorded dataset cannot be used to perform a shape analysis of the Michel spectrum. This is because the NP final state only contains a single positron and would thus not pass event filtering on the online GPU farm~\cite{Henkys:2022ste}. A dedicated search strategy can however be envisaged, {\emph{e.g.}} by performing an analysis on histograms which are filled with track information as part of the online track reconstruction~\cite{Perrevoort:2018okj,Mu3e:2020wph,Henkys:2022ste}. This strategy has been proposed by the Mu3e collaboration to look for LFV two-body decay $\mu^+ \to e^+ \phi$ with $\phi$ going undetected~\cite{Perrevoort:2018okj,Perrevoort:2018ttp,Hesketh:2022wgw}. This also allows for a larger dataset, of order of ${\cal O}(10^{15})$  muons, at the price  of a worse positron energy resolution. We anticipate that, given the large size of the dataset, the proposed Mu3e analysis will be dominated by systematic uncertainties, as is already the case for MEG II with a dataset smaller by several orders of magnitude. For this reason in the Mu3e analysis we will fix
keep fixed the Michel spectrum normalization to $N_{\mu^+}=10^{10}$ and vary now the systematic uncertainty in the range $\epsilon_{\rm syst}=10^{-3}\,-\,  10^{-2}$. Although Mu3e does not quote a global systematic uncertainty, detector-related uncertainties (tracking efficiency, alignment, momentum scale) are individually controlled at the sub-percent level~\cite{Mu3e:2020gyw}. As for the case of MEG-II, we first show the limit on the parameter $\alpha$ of Eq.~\eqref{eq:alpha} as a function of the DM mass in the bottom panels of  Fig.~\ref{fig:alphabounds} for the case of vector bilinear and for the same assumptions on the interaction coupling structure and binning choices as before. Also here we assume that the region close to the Michel spectrum endpoint might be used for calibration purposes, implying than our limits cannot be trusted for $m_\chi < 1\,$MeV. We see that, thanks to the lower threshold for positron energies, the bound can extend to larger masses. Again, we translate the limit on $\alpha$ onto bounds on the NP scale $\Lambda$ by using Eq.~\eqref{eq:alpha} and show the corresponding results in the right panel of Fig.~\ref{fig:Bound_inv_search}.
Here, we present the results by fixing the number of muons to be $10^{10}$ and varying the systematic uncertainties from $10^{-3}$ to $10^{-2}$ (indicated by the shaded bands, with the central value shown as a solid line). The constraints are displayed for the $X=R$ scenario in blue, the $X=L$ scenario in orange and $X=V,A$ case in purple. The results show that for the RH combination the bound is flat and of $\mathcal{O}(\mathrm{2~TeV})$, while for the LH combination the bound becomes weaker at lower masses, since in the limit $m_\chi \to 0$ the NP signal is completely degenerate with the SM one. Additionally, compared to MEG-II, the constraints extend to higher masses due to the positron energy efficiency Mu3e, which starts operating at lower positron energies ($\sim 15~\mathrm{MeV}$) with respect to the energy efficiency of MEG-II ($\sim 40~\mathrm{MeV}$).

As in the MEG-II analysis, the Mu3e bounds exhibit the same qualitative behavior. At small $m_\chi$, the sensitivity is dominated by the angular information, with the $X=R$ scenario yielding a stronger constraint than the $X=V,A$ cases. The sensitivity is maximal around $m_\chi\sim\mathcal{O}(30~\mathrm{MeV})$, where both angular and energy information contribute, while at larger masses the different scenarios lead to comparable bounds up to the kinematic endpoint.

\section{Radiative decay at MEG-II}\label{sec:mutoegamma}

In Sec.~\ref{sec:inv_decay} we have investigated how the MEG-II and Mu3e datasets allow us to constrain the LFV interactions of Eq.~\eqref{eq:lag} by a shape analysis of the Michel spectrum, testing NP scales of the order of a few TeV. The major drawback of that analysis is that there is no sensitivity to DM masses below $\simeq 1\,$MeV. This is because in this case the NP distortion to the Michel spectrum lies too close to the distribution endpoint $E_e = m_\mu/2$, a region that is generally assumed to be signal-free and used for calibration (this is not anymore true for the scalar and tensor interactions for which, also in the massless limit, the distributions are different from the SM one as shown in App.\,\ref{app:distr_tensor}).
Nevertheless, in the vector scenario one can take advantage of the excellent MEG-II photon reconstruction to exploit the radiative decay $\mu^+ \to e^+ \chi \bar \chi \gamma$ and perform a shape analysis against the SM RMD $\mu^+ \to e^+ \bar \nu_\mu \nu_e \gamma$. Having calibrated the experiment on the three-body $\mu^+$ decay, this process can then be used to constrain DM lighter than the one probed by the invisible decay searches proposed for MEG-II in Sec.~\ref{sec:MEG-II_inv}. This strategy has been recently proposed in the context of MEG-II to search for invisible axions in LFV decays $\mu^+ \to e^+ a \gamma$~\cite{Jho:2022snj}.

The MEG-II trigger and analysis are optimised to search for the two-body decay $\mu^+ \to e^+\gamma$, where the positron and the photon are emitted back to back, both carrying an energy of about a half of the muon mass. As a consequence, the experiment is clearly not optimised to search for the process $\mu \to e \bar{\chi}\chi\gamma$, a four-body decay with very different kinematics. Nevertheless we show that, without any change to the experimental analysis strategy and without the need of a dedicated MEG-II run, one is able to set relevant bounds on the interactions of Eq.~\eqref{eq:lag}, extending the results on the previous section to lower DM masses.

We adopt a similar approach to the one employed for the invisible $\mu^+ \to e^+\chi\bar \chi$ search and perform a shape analysis on the RMD spectrum, where one now has access to a larger number of kinematic variables. In particular we choose the positron and photon energies $E_e$ and $E_\gamma$ and the polar and azimuthal angles between them\,\footnote{Note that MEG-II defines 
$\theta_{e\gamma} = \pi - \theta_e - \theta_\gamma$ and 
$\phi_{e\gamma} = \pi + \phi_e - \phi_\gamma$ so that the back to back topology corresponds to $\theta_{e\gamma}=\phi_{e\gamma}=0$
} $\theta_{e\gamma}$ and $\phi_{e\gamma}$.
We follow the analysis of Ref.~\cite{Jho:2022snj} and model the differential efficiency for the MEG-II trigger by assuming that it factorizes as
\begin{equation}\label{eq:trigger-rad}
{\cal A}_{\rm trig}^{\rm MEG-II}= 
{\cal A}_e(E_e)\times 
{\cal A}_\gamma(E_\gamma)\times 
{\cal A}_{\theta_{e\gamma}}(E_e,\theta_{e\gamma}) \ ,
\end{equation}
and define the fiducial region $\fid$ with the following requirements energy and geometrical acceptance requirements: $E_e>45\,$MeV, $E_\gamma>40\,$MeV, $|\cos\theta_e|<0.35$, $|\cos\theta_\gamma|<0.35$, $\frac{2\pi}{3}<\phi_e<\frac{4\pi}{3}$, $|\phi_\gamma|<\frac{\pi}{3}$, $|\theta_{e\gamma}|<0.3$ and $|\phi_{e\gamma}|<0.3$.  Differently from Sec.~\ref{sec:inv_decay} where we have used analytical expression for the $1\to 3$ SM and NP $\mu^+$ decay, here we rely on a Monte-Carlo simulation performed with {\tt  MadGraph5\_aMC@NLO}~\cite{Alwall:2014hca} after having implemented the interactions of Eq.~\eqref{eq:lag} in the {\tt Feynrules} package~\cite{Alloul:2013bka} using the {\tt UFO} format~\cite{Degrande:2011ua}. Within this setup and restricting to the fiducial region $\fid$ defined above, we find a SM RMD branching ratio of ${\rm BR}^\fid_{\rm RMD}=2.8 \times 10^{-9}$, yielding a total number of events in the fiducial region of 
$N^\fid_{\rm RMD} = N_{\mu^+} \times \mathrm{BR}^\fid_{\rm RMD} \times \epsilon_{\rm trig}^{\rm MEG-II} \simeq 2\times10^4$, computed using a total number of muons $N_{\mu^+}=10^{14}$. This estimate is compatible with the measure of the RMD at MEG~\cite{MEG:2013mmu}, and consistent with analytic computation~\cite{Fischer:1994pn, Kuno:1999jp, Fronsdal:1959zzb, Behrends:1955mb, Lenard:1953zz}.
These backgrounds are characterized by a small time separation $\Delta t_{e\gamma}$ between the positron and the photon.   

As mentioned in Sec.~\ref{sec:MEG-II_inv}, on top of RMD events, MEG-II also observes RC background events where a photon originating either from a RMD or from photon pair production in the detector material is paired up with a Michel positron. These RC backgrounds have larger value of $\Delta t_{e\gamma}$ with respect to the RMD one. Here we only consider the first of these two sources of RC background and estimate it as follows, closely following Ref.~\cite{Jho:2022snj}. We simulate RMD events where a positron has has $E_e < 40\,$MeV and is therefore not reconstructed\,\footnote{Strictly speaking, there is a non-zero probability of missing a positron also for $E_e>40\,$MeV, due to the not ideal efficiency for positron reconstruction. We however find this contribution to be marginal.} or it is outside the angular detector geometrical acceptance of $\fid$, while requiring the photon to be within $\fid$. Then we generate a Michel sample with $E_e>45~{\rm MeV}$, requiring the positron to be within the $\fid$ geometrical acceptance. Finally, we build the RC sample by randomly associating to each Michel event a photon from the RMD sample, further imposing the $\theta_{e\gamma}$ and $\phi_{e\gamma}$ requirements. With this procedure the effective BR for the RC background within $\fid$ can be expressed as~\cite{Jho:2022snj}
\begin{equation}
{\rm BR}_{\rm RC}^\fid={\rm BR}_{\rm Mich} \times {\rm BR}_{\rm RMD'}\times (R_\mu^+ \times \Delta t_{e\gamma}) \ ,
\end{equation}
where ${\rm BR}_{\rm Mich}$ and ${\rm BR}_{\rm RMD'}$ are the Michel and RMD BRs computed as described above
and the last factor represents the probability that a photon from a RMD is accidentally associated with a Michel positron within a time window $\Delta t_{e\gamma}$. Here $R_{\mu}^+=3\times 10^7/\rm{s}$ and for the trigger window of $\Delta t_{e\gamma}=24\,{\rm ns}$ this probability is approximately $72\%$, leading to ${\rm BR}_{\rm RC}=5.1\times10^{-8}$. This is larger than the RMD estimate by about one order of magnitude, confirming that RC is the dominant background source for MEG-II (as anticipated in Sec.~\ref{sec:MEG-II_inv}). In view of this result for our sensitivity estimate we only retain this source of background neglecting the RMD one. Moreover this leads to a number of RC background events of 
 $N^\fid_{\rm RC}  \simeq 10^5$ after having taken into account the trigger efficiency of Eq.~\eqref{eq:trigger-rad}, which motivates our choice of rate normalization for the invisible decay search of Sec.~\ref{sec:MEG-II_inv}. 
 
 As for the case of the invisible search of Sec.~\ref{sec:inv_decay}, we build a $\chi^2$ distribution with six equally spaced bins in the variables $E_e$\,, $E_\gamma$,\, $\theta_{e\gamma}$\, and $\phi_{e\gamma}$, taking $N_{\mu^+}=10^{14}$~\cite{Baldini:2023hxt} and fixing the experimental systematic uncertainty to $\varepsilon_{\rm syst}=4\times 10^{-2}$, which is dominated by the photon trigger uncertainty. As for the invisible search, the number of bins is chosen in accordance with the experiment’s resolution to capture the key features of the distribution.
Our results are illustrated in Fig.~\ref{fig:RMDbound}, where the bounds obtained from the radiative-decay analysis are shown as solid lines and compared with those derived from the invisible search, shown as dashed lines. Results are presented for the pure RH scenario (blue), the pure LH scenario (orange), and the pure vector or axial scenarios (purple), with solid blue and orange curves overlapped. Since the radiative analysis was performed only for the $X=R,L$ scenarios, no solid purple curve is shown. For the invisible search, we assume a permille level systematic uncertainty and perform the analysis using $N_{\mu^+}=10^5$, 10×10 binning in positron energy and $\cos\theta_e$. In contrast, for the radiative analysis we fix the systematic uncertainty at 4\%, including an additional contribution from the photon trigger uncertainty, and use six bins for each of the kinematic variables $E_e$, $E_\gamma$, $\theta_{e\gamma}$, and $\phi_{e\gamma}$. The key result is that, due to the different kinematics, the radiative analysis extends the constraints into the shaded gray region corresponding to the calibration region. Moreover, unlike the invisible search, where the pure LH case becomes degenerate with the Michel spectrum in the massless limit, the $\mu \to e \chi \chi \gamma$ channel remains sensitive to $\mathcal{O}(\mathrm{TeV})$ scales even in the $m_\chi \to 0$ limit. In the radiative decay analysis, we did not consider the pure vector and axial scenarios separately, as their expected behavior is similar to that of the pure RH/LH cases. The results for the scalar and tensor interaction are shown in Fig.~\ref{fig:boundMEGII_RMD_tensor_scalar} in App.\,\ref{app:distr_tensor}.

\begin{figure}
    \centering
\includegraphics[width=0.5\linewidth]{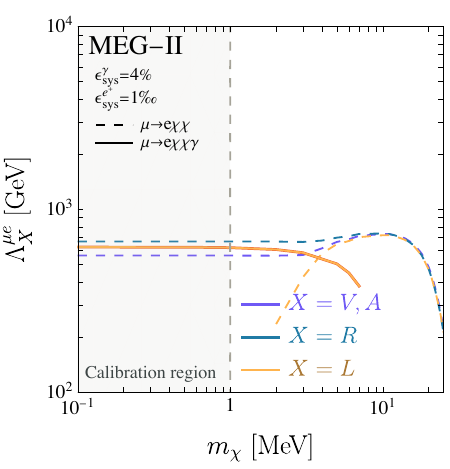}
    \caption{95\% CL constraints  on $\Lambda^{\mu e}_X$ as a function of the DM mass for $X=R$ (blue lines) and $X=L$ (orange lines) and $X=V,A$ (purple line). The dashed lines correspond to the results shown in the left panel of Fig.~\ref{fig:Bound_inv_search}, obtained using $N_{\mu^+}=10^5$ and 10×10 bins in positron energy and $\cos\theta_e$. The solid lines correspond to the results of the radiative analysis, performed with $6^4$ bins for ${E_e, E_\gamma, \theta_{e\gamma}, \phi_{e\gamma}}$. The solid blue and orange lines are overlapped. Additionally, we assume a systematic uncertainty at the permille level, while for the radiative analysis we fix the systematic uncertainty at 4\%.}
    \label{fig:RMDbound}
\end{figure}

\section{Complementary bounds}\label{sec:complementary}
 Having discussed the direct laboratory limits on the LFV operators in Eq.~\eqref{eq:lag}, we now turn to complementary constraints from other laboratory experiments, astrophysical observations, and cosmological probes. These constraints can either test the same flavor-violating interactions through different observables or probe additional directions in flavor space beyond the single off-diagonal combination accessible to the LFV searches discussed in Secs.~\ref{sec:inv_decay} and~\ref{sec:mutoegamma}. In particular, they can constrain flavor-diagonal interactions involving electrons or muons, which do not contribute to the MEG-II or Mu3e processes considered above, but may naturally be generated together with the LFV operators in realistic UV completions. Their implications are therefore relevant for assessing the overall phenomenological viability of the EFT.

Using the same Lorentz basis and conventions as in Eq.~\eqref{eq:lag}, we parametrize the flavor-diagonal interactions as
\begin{equation}
\label{eq:FC}
\mathcal{L}_{\rm diag}
=
\frac{1}{\Lambda^2}
\left[\sum_{X}
\left(
\bar{e}\,c_{X}^{ee}\Gamma_X e
\right)
\left(
\bar{\chi}\,c_{X}^{\chi}\Gamma_X\chi
\right)+\left(
\bar{\mu}\,c_{X}^{\mu\mu}\Gamma_X\mu
\right)
\left(
\bar{\chi}\,c_{X}^{\chi}\Gamma_X\chi
\right)\right] + \text{h.c.}
\,.
\end{equation}
As before, we focus on scenarios in which the DM and leptonic sectors share the same bilinear structure, i.e.\ $X=Y$. Accordingly, we define the effective scale associated with flavor-diagonal interactions as $\Lambda^{\ell\ell}_{X}\equiv
\Lambda/\sqrt{c_{X}^{\ell\ell}c_{X}^{\chi}}$, with $\ell=e,\mu$, where $c_{X}^{\ell\ell}$ denotes the Wilson coefficient of the corresponding flavor-diagonal lepton bilinear. Although one may naturally expect $\Lambda^{\mu e}_{X}\lesssim\Lambda^{\mu\mu,ee}_{X}$, in the following we treat flavor-conserving and flavor-violating couplings as independent EFT parameters and simply compare the corresponding phenomenological constraints, without assuming a specific flavor structure. In Appendix~\ref{App:FSE}, we briefly discuss a possible flavor scheme in which
the operator inducing $\mu\to e\chi\bar\chi$ can remain sizable, while other flavor-related operators are sufficiently suppressed to satisfy stringent laboratory bounds from processes such as $\mu\to3e$. This construction is not intended as a complete UV model, but rather as a proof of principle that the required flavor hierarchy can be consistently realized. As a shorthand, we denote the electron- and muon-flavor-diagonal interactions by $\mathcal{O}_{ee\chi\chi}$ and $\mathcal{O}_{\mu\mu\chi\chi}$, corresponding to the first and second terms of Eq.~\eqref{eq:FC}, respectively. In each case, we specify the relevant Lorentz structure and coupling assignment.

We organize the discussion as follows. We first consider constraints on flavor-conserving interactions and then discuss complementary bounds on the same off-diagonal interactions probed by the LFV searches.

\subsection{Flavor-diagonal constraints}
\paragraph{Bounds from DM Direct Detection:} 
The flavor-diagonal electron operator $\mathcal{O}_{ee\chi\chi}$ 
can in principle be probed by DD experiments through DM elastic scattering on atomic electrons, $\chi e \to \chi e$, which is the most relevant channel in the light DM regime. Indeed, for $m_\chi$ in the keV--MeV range, conventional nuclear recoil searches rapidly lose sensitivity, since the kinetic energy transferred to nuclei falls below typical experimental thresholds. The appropriate probes are instead low threshold searches based on electron ionization, electron excitation, absorption, or, in more specialized scenarios, collective excitations such as magnons or phonons~\cite{Essig:2011nj,Hochberg:2015pha,Hochberg:2017wce,Knapen:2017xzo,Trickle:2019nya,Trickle:2019ovy,Knapen:2017ekk,Mitridate:2021ctr,Derenzo:2016fse,Essig:2015cda,Caputo:2019cyg,Berghaus:2022pbu}.

In the MeV mass range, the leading DD constraints arise from electron recoil searches. Current bounds on spin independent DM--electron scattering are provided by low threshold semiconductor experiments based on skipper CCDs, such as SENSEI~\cite{SENSEI:2023zdf} and DAMIC-M~\cite{DAMIC-M:2023gxo,DAMIC-M:2025luv}, together with cryogenic detectors such as SuperCDMS~\cite{SuperCDMS:2024yiv} and noble liquid S2-only analyses from XENON10~\cite{Essig:2012yx} and DarkSide-50~\cite{DarkSide:2018ppu}.

Among the possible four-fermion structures, $VV$ and
$SS$ interactions generate unsuppressed non-relativistic
$\chi e$ scattering amplitudes. Other Lorentz structures instead lead to parametrically suppressed rates in the non-relativistic regime. For definiteness, we focus on the $VV$ interaction, for which the DM--electron scattering cross section is
\begin{equation}
\sigma_{\chi e} =
\frac{(c^{ee}_V  c_V^\chi)^2 }{\pi \Lambda^4} \mu_{\chi e}^2
\simeq
3.2\times10^{-47}~{\rm cm}^2 
\left(\frac{1~{\rm TeV}}{\Lambda_{V}^{ee}}\right)^4
\left(\frac{\mu_{\chi e}}{m_e}\right)^2 ,
\end{equation}
where $\mu_{\chi e}$ is the reduced mass of the $\chi$--$e$ system. 
For a heavy mediator and a DM mass in the $1-30~{\rm MeV}$ interval, the strongest present limits from SENSEI, DAMIC-M, and DarkSide-50 are typically in the range $10^{-34} - 10^{-38}~{\rm cm}^2$. Using the estimate above, and taking $\mu_{\chi e}\simeq m_e$, this corresponds only to $\Lambda_{V}^{ee} \simeq 0.8 - 8~{\rm GeV}$. Equivalently, a TeV-scale contact interaction would predict $\sigma_{\chi e}\sim 10^{-47}~{\rm cm}^2$, far below the current reach of electron recoil DD searches. Therefore, although DD provides a clean probe of the flavor-diagonal electron interaction, present bounds are not competitive with the energy scales tested by LFV observables in the parameter space considered here.

Future experiments exploiting collective excitations in superconductors~\cite{Hochberg:2015pha}, polar materials~\cite{Knapen:2017xzo,Griffin:2018bjn,Baiocco:2025omh}, Dirac materials~\cite{Hochberg:2017wce}, and phonon-based detectors~\cite{Coskuner:2021qxo,SuperCDMS:2020aus} are expected to extend the sensitivity to lighter DM masses in the sub-MeV and keV range. Even under optimistic projections, however, the corresponding reach in the effective scale $\Lambda_{X}^{ee}$
 remains significantly below the scales probed by LFV searches. 

\paragraph{Bounds from Indirect Detection:}
A complementary probe of the flavor-diagonal electron operator
$\mathcal{O}_{ee\chi\chi}$
is provided by ID searches for
DM annihilation into charged leptons, $\chi \bar\chi\to e^+e^-$. In the
MeV$-$GeV mass range, the most relevant constraints arise from energy
injection at recombination, constrained by CMB anisotropies, and from
late time Galactic signals. These include low-energy cosmic ray
electrons and positrons measured by Voyager~1 and AMS-02, diffuse
X-ray and gamma-ray observations by INTEGRAL, XMM-Newton, NuSTAR and Fermi-LAT, as well as searches for the Galactic 511~keV line
produced by positron annihilation in the interstellar medium
\cite{Planck:2018vyg, Slatyer:2015jla,Cirelli:2009bb,Cirelli:2009vg, Boudaud:2016mos,Boudaud:2018hqb,Cirelli:2010xx,Cirelli:2020bpc,
Cirelli:2023tnx,Siegert:2024hmr,Fermi-LAT:2009ppq,Fermi-LAT:2010cni,
Hooper:2010mq,Boehm:2003bt,Beacom:2004pe,Sizun:2006uh,
DelaTorreLuque:2023cef,Aghaie:2025dgl}.

\smallskip
For annihilation, different Lorentz structures also lead to different velocity and chirality suppressions. However, the situation is less restrictive than in non-relativistic direct detection, since the annihilation rate is controlled by the hard scale $m_\chi$, rather than by the small momentum transfer or halo velocity relevant for elastic scattering. The vector current--current interaction can therefore be used as a representative benchmark, keeping in mind that the precise indirect-detection limits retain a mild dependence on the operator structure. In the limit $m_\chi\gg m_e$, one obtains
\begin{equation}
\langle \sigma v_{\rm rel}\rangle_{\chi\chi \to ee} =
\frac{(c^{ee}_V  c_V^\chi)^2 }{\pi \Lambda^4} m_\chi^2
\simeq
3.8\times10^{-34}~{\rm cm^3\,s^{-1}} \,  
\left(\frac{1~{\rm TeV}}{\Lambda^{ee}_{V}}\right)^4
\left(\frac{m_\chi}{10~{\rm MeV}}\right)^2 \ .
\end{equation}

For MeV-scale DM, the strongest generic constraints on the annihilation rate arise from CMB energy injection, which typically requires $\langle\sigma v_{\rm rel}\rangle \lesssim 10^{-30} – 10^{-28}~{\rm cm^3 \, s^{-1}}$, depending on the DM mass and on the efficiency factor. For a TeV--scale contact interaction with an $\mathcal O(1)$ Wilson coefficient, the annihilation rate is much smaller that current limits in the region of $\Lambda$ probed by LFV searches. ID constraints are therefore safely negligible. For $m_\chi < m_e$, the tree level annihilation channel
$\chi\bar\chi\to e^+e^-$ is instead kinematically forbidden. 
The remaining loop induced annihilation channels into photons are highly suppressed and completely negligible for the parameter space considered here.

\paragraph{Bound from Supernova Cooling:}
Proto-neutron stars (PNS) formed in core-collapse supernovae provide an efficient environment for the production of light dark sector particles. If such particles are sufficiently long-lived, they can freely stream out of the stellar medium and contribute to supernova cooling, leading to constraints from the observed neutrino signal of SN1987A~\cite{Raffelt:1996wa}. Recently the resulting limits have been investigated in Ref.~\cite{Manzari:2023gkt}   for dark fermions coupled to SM leptons through flavor-diagonal four-fermion operators involving electrons and muons, namely $\mathcal{O}_{ee\chi\chi}$ and $\mathcal{O}_{\mu\mu\chi\chi}$ (for earlier studies of dark fermions see e.g. Refs.~\cite{Sutherland:1975dr, Dicus:1976ra, Raffelt:1987yt, Barbieri:1988av, Mohapatra:1990vq, Dreiner:2003wh, Dreiner:2013mua, Guha:2018mli}). For dark-fermion masses below $\mathcal{O}(100 \,\mathrm{MeV})$, the resulting cooling constraints probe effective interaction scales of several TeV, as shown in the second ($\Lambda^{\mu\mu}_{X}$) and third ($\Lambda^{ee}_{X}$) columns of Tab.\,\ref{tab:SNlimits}, taken from Ref.~\cite{Manzari:2023gkt}.

\smallskip
At large $\Lambda^{\ell\ell}_{X}$, production rates are too small to appreciably affect the neutrino signal. Conversely, at small $\Lambda^{\ell\ell}_{X}$, dark fermions become trapped and are effectively emitted thermally from the surface of a ``dark sphere'', similarly to neutrinos. For sufficiently strong interactions, this sphere becomes larger and hence colder than the neutrino sphere, so that dark radiation ceases to act as an efficient energy-loss channel. Accordingly, the lower edge of each interval in Tab.\,\ref{tab:SNlimits} is determined by trapping, while the upper edge is set by free-streaming. The bounds were obtained for $m_\chi = 0$ and remain approximately valid as long as dark fermions can be efficiently produced in the PNS core. In practice, this corresponds to $m_\chi \lesssim 100\,\MeV$ in the free-streaming regime and $m_\chi \lesssim 10\,\MeV$ in the trapping regime.

As we discuss in the following, the trapping regions at low values of $\Lambda^{\ell\ell}_{X}$ are essentially excluded by direct experimental searches. 
Consequently, the SN1987A cooling argument excludes effective scales for both electron and muon couplings up to $\Lambda^{ee,\mu\mu}_{X} \simeq 5\text{--}7\,\TeV$, depending on the Lorentz structure. These limits are therefore among the strongest probes of flavor-diagonal EFT operators in the light DM mass range considered here.

\begin{table}[t]
\renewcommand{\arraystretch}{1.6}
  \setlength{\arrayrulewidth}{.35mm}
\setlength{\tabcolsep}{3.9 mm}
\centering
\begin{tabular}{|c||c|c||c|}
\hline 
$X$ & $\Lambda^{\mu \mu}_{X}$ [TeV] & $\Lambda^{ee}_{X}$ [TeV] &$\Lambda^{\mu e}_{X}$ [TeV]\\
\hline 
$S$ & $0.0017 - 4.4$  & $0.070 - 5.4$ & $2.0$\\
$V$ & $0.0017 - 5.7$ & $0.11 - 5.9$& $2.8$\\
$T$ & $0.0033 - 6.8$   & $0.17 - 7.0$& $4.4$ \\
\hline
\end{tabular}
\caption{SN~1987A exclusion regions for the effective scale of both flavor-diagonal and flavor off-diagonal operators. For flavor-diagonal interactions, the limits are taken from Ref.~\cite{Manzari:2023gkt}. We show only the scalar-scalar, vector-vector, and tensor-tensor Lorentz structures, assuming $m_\chi \simeq 0$. For these flavor-diagonal operators, the lower boundary of each excluded interval is determined by the trapping regime, while the upper boundary is set by the free-streaming regime.} 
\label{tab:SNlimits}
\end{table}

\paragraph{Bounds from colliders and fixed-target experiments}
We conclude this subsection by discussing complementary laboratory bounds on flavor-diagonal operators from collider searches and fixed target experiments. The operator $\mathcal{O}_{ee\chi\chi}$ is directly probed at LEP through mono-photon events, $e^+e^-\to\chi\bar\chi\gamma$, where an initial state photon recoils against missing energy. The relevant experimental input is provided by the LEP single photon and missing energy searches, in particular the DELPHI analyses~\cite{DELPHI:2003dlq,DELPHI:2008uka}, which were
recast in terms of effective DM--electron contact operators in
Ref.~\cite{Fox:2011fx}. For light DM and Wilson coefficients of order unity, these searches constrain the EFT cutoff at the level of a few hundred GeV, typically $\Lambda^{ee}_{X} \gtrsim \mathcal{O}(500 \GeV)$.

Muon fixed target experiments can instead probe the diagonal muon operator
$\mathcal{O}_{\mu\mu\chi\chi}$. In particular, the NA64 collaboration has searched for invisible
particles produced by a $160~{\rm GeV}$ muon beam impinging on a fixed target, using a missing-energy/momentum signature~\cite{NA64:2024klw}. The relevant process for the contact operator is $\mu N\to\mu N\chi\bar\chi$, which 
for vector interactions was derived in Ref.~\cite{Crivelli:2025pjb}, yielding $\Lambda^{\mu \mu}_{X}\gtrsim\mathcal{O}(10 \GeV)$.

 \subsection{Flavor off-diagonal constraints}
Apart from constraints on the total muon lifetime as part of the EW fit, the LFV operators in Eq.~\eqref{eq:lag} are only constrained by supernova cooling. It is well known that a sizable population of muons exist in the PNS plasma, which has been used to constrain flavor-conserving muon couplings of dark particles produced from muon initiated scattering processes~\cite{Bollig:2020xdr,Croon:2020lrf, Caputo:2021rux,Manzari:2023gkt}. As originally proposed in Ref.~\cite{MartinCamalich:2020dfe} for hyperons, the same analysis can be applied to two-body muon decays, where the simple kinematics allow for a analytic estimate of the energy loss rate, see Refs.~\cite{Calibbi:2020jvd, MartinCamalich:2025srw, Ziegler:2026kis}. Here we use the same estimate for the three-body decay to dark fermions, leaving a complete numerical analysis for future research. Since the energy fraction carried away by dark particles in the three-body decay is larger than in the two-body decay, this estimate should be rather conservative. 

In the non-relativistic limit and neglecting final state Pauli blocking and the dark fermion mass, the energy loss rate  $Q$  per unit volume  can be estimated very roughly as the product of {\emph{(i)}} the muon decay rate, {\emph{(ii)}} the  energy released per decay, approximately given by the muon mass, and {\emph{(iii)}} the muon number density $n_\mu$. The latter may be estimated as  the electron number density $n_e$ times a Boltzmann factor, giving in total
\begin{align}
Q_{\mu \to e \chi \overline \chi} \simeq  n_e m_\mu \Gamma(\mu\to e  \chi \overline \chi) \exp \left(- \frac{m_\mu - m_e}{T}\right) \, . 
\end{align}
Dividing by the nuclear mass density, $\rho \simeq m_n n_n$, gives the emissivity
\begin{equation}
\epsilon_{\mu \to e \chi \overline \chi} \simeq   \frac{n_e m_\mu}{n_n m_n} \frac{{\rm BR} (\mu\to e  \chi \overline \chi )}{\tau_\mu}  \exp \left(- \frac{m_\mu - m_e}{T}\right) \simeq \frac{10^{19}  {\rm erg}}{{\rm g \, s}}  \left(\frac{{\rm BR}(\mu \to e \chi \overline \chi)}{6 \times 10^{-5}} \right) \, ,
\end{equation}
where, to obtain the approximate numerical result, we have used characteristic values for a PNS temperature of $30 \MeV$ and an electron fraction of $Y_e = n_e/n_n \approx 0.12$~\cite{Garching}. Using instead the muon number fraction directly, $Y_\mu \approx 0.026$~\cite{Garching}, increases the emissivity by a factor of about 7. The former estimate is, however, in better agreement with the analysis of Ref.~\cite{Zhang:2023vva}, which considered dark-boson production from $\mu$--$p$ bremsstrahlung induced by the same LFV coupling and found ${\rm BR} (\mu \to eX)\lesssim 9 \times 10^{-6}$. Finally imposing ${\rm BR} (\mu \to e \chi \overline \chi)\lesssim 6 \times 10^{-5}$ yields the limits
\begin{align}
\Lambda_{S}^{\mu e} & \gtrsim 2.0 \TeV \, , & 
\Lambda_{V}^{\mu e} & \gtrsim 2.8 \TeV \, , & 
\Lambda_{T}^{\mu e} & \gtrsim 4.4 \TeV \, ,
\end{align}
which are of the same order as the SN~1987A constraints on flavor-diagonal interactions. Using the muon number fraction directly would reduce these limits by about 40\%. We emphasize that these numbers are only rough estimates and should ultimately be superseded by a dedicated analysis including the full radial profiles, relativistic corrections, Pauli blocking and possible medium effects, similar to the analysis performed in Ref.~\cite{Camalich:2020wac} for hyperon decays.

\section{Thermal production}\label{sec: Thermal production}
In Sections~\ref{sec:inv_decay} and \ref{sec:mutoegamma} we have assessed the sensitivity of high-intensity muon factories to the LFV interactions in Eq.~\eqref{eq:lag}, which reach UV scales of order TeV. We now turn to  thermal production of dark fermions in the early universe and investigate whether the same LFV interactions can account for the observed DM relic abundance.

We assume that DM is produced via the {\it freeze-in} mechanism~\cite{Hall:2009bx} through LFV processes involving muons and electrons in thermal equilibrium with the SM plasma at temperature $T$ (see also Ref.~\cite{Belfatto:2021ats}). The off-diagonal operators in Eq.~\eqref{eq:lag} give rise to two production channels: the three-body decay $\mu\to e\chi\bar\chi$ and the annihilation process $\mu\bar e\to\chi\bar\chi$. We neglect the crossed process $\mu\chi\leftrightarrow e\chi$, since during freeze-in the DM abundance remains much smaller than its equilibrium value. The decay contribution is IR dominated and therefore independent of the reheating temperature, provided that $T_R$ is sufficiently high, whereas the annihilation contribution is UV sensitive and depends explicitly on $T_R$.

Assuming a negligible initial DM abundance at reheating and neglecting inverse
processes, as appropriate in the freeze-in regime, the Boltzmann equation in
terms of the dimensionless parameter $x=m_\mu/T$ reads
\begin{equation}\label{eq:boltzmann}
\frac{{\rm d}Y_\chi}{{\rm d}x}
\simeq
-
\frac{\sum_i {\cal C}_i(x)}
{\mathcal S H x}
\left(
\frac{1}{3}
\frac{{\rm d}\ln\mathcal S}{{\rm d}\ln x}
\right)\,.
\end{equation}
Here, the sum includes the decay and annihilation contributions, namely
$\sum_i{\cal C}_i(x)\equiv{\cal C}_{1\to3}(x)+{\cal C}_{2\to2}(x)$.
Moreover, $Y_\chi=n_\chi/\mathcal S$ is the yield of $\chi$ particles, with
$n_\chi$ their number density and
$\mathcal S=(2\pi^2/45)\,g_{*s}(x)m_\mu^3/x^3$ the entropy density. During
radiation domination, the Hubble expansion rate is
$H=\sqrt{8\pi^3g_*(x)/90}\,m_\mu^2/(M_{\rm Pl}x^2)$, where
$M_{\rm Pl}\simeq1.22\times10^{19}\,{\rm GeV}$. The quantities $g_{*s}$ and
$g_*$ denote the effective numbers of relativistic degrees of freedom
contributing to the entropy and energy densities, respectively. The collision
terms entering Eq.~\eqref{eq:boltzmann} are
\begin{equation}\label{eq:collision}
\begin{split}
{\cal C}_{1\to3}(x)
\equiv
2{\cal C}_{\mu^-\to e^-\chi\bar\chi}(x)
&=
2\frac{m_\mu^3}{\pi^2 x}
K_1\left(x\right)
\Gamma_{\mu\to e\chi\bar\chi}\,,
\\
{\cal C}_{2\to2}(x)
\equiv
2{\cal C}_{\mu^-\!e^+\to\chi\bar\chi}(x)
&=
2\frac{m_\mu^6}{4\pi^4 x}
\int_1^{\infty}{\rm d}y\,
\left(y^2-1\right)^2
K_1\left(xy\right)
\sigma_{\mu\bar e\to\chi\bar\chi}(y)\,.
\end{split}
\end{equation}
Here, $y=\sqrt{s}/m_\mu$, with
$s=(p_\mu+p_e)^2$ the squared center-of-mass energy, while $K_1$ is the
modified Bessel function of the second kind of order one. The factors of $2$
in Eq.~\eqref{eq:collision} account for the charge-conjugate processes
$\mu^+\to e^+\chi\bar\chi$ and
$\mu^+e^-\to\chi\bar\chi$, respectively. Since $Y_\chi$ tracks the abundance
of $\chi$ particles only, no additional factor of $2$ associated with the
simultaneous production of $\bar\chi$ is included in the collision terms. In deriving Eq.~\eqref{eq:collision}, we approximate the Fermi--Dirac distributions by Maxwell--Boltzmann distributions, neglect Pauli-blocking
effects, and set the electron mass to zero. The lower integration limit in
the scattering contribution is $y=1$, since the decay channel considered
here is open only for $m_\chi<m_\mu/2$. At the reheating temperature $T=T_R$, corresponding to
$x_R=m_\mu/T_R$, we impose the initial condition $Y_\chi(T_R)=0$. The
present-day yield, $Y_\chi^0\equiv Y_\chi(T_0)$, where
$T_0\simeq2.73\,{\rm K}$ is the present CMB temperature, is obtained by
integrating Eq.~\eqref{eq:boltzmann} from reheating to the present epoch.
Since DM production becomes negligible at temperatures well below the muon
mass, the lower temperature limit can safely be set to zero. By contrast, the
upper limit $T_R$ must be retained because the annihilation contribution is
UV sensitive. The present-day yield can therefore be written as
\begin{equation}\label{eq:freeze-in-integral}
Y_\chi^0
\simeq
1.4\,M_{\rm Pl}
\sum_i{\cal I}_i(x_R)\,,
\qquad
{\cal I}_i(x_R)
\equiv
\frac{1}{m_\mu^5}
\int_{x_R}^\infty{\rm d}x\,
\frac{x^4{\cal C}_i(x)}
{g_{*s}\sqrt{g_*}}
\left(
1-\frac{x}{3g_{*s}}
\frac{{\rm d}g_{*s}}{{\rm d}x}
\right)\,,
\end{equation}
where, analogously to the collision terms,
$\sum_i{\cal I}_i(x_R)\equiv{\cal I}_{1\to3}(x_R)+{\cal I}_{2\to2}(x_R)$.
The corresponding total relic abundance, including both $\chi$ and
$\bar\chi$, is
\begin{equation}\label{eq:relic}
\Omega_\chi h^2
=
2\,
\frac{m_\chi\mathcal S_0}
{\rho_c^0/h^2}
Y_\chi^0
\simeq
2.8\,
\frac{M_{\rm Pl}\mathcal S_0}
{\rho_c^0/h^2}\,
m_\chi
\sum_i{\cal I}_i(x_R)\,.
\end{equation}
The factor of $2$ accounts for the equal abundance of antiparticles produced
by the same processes, $Y_{\bar\chi}=Y_\chi$. If $\chi$ constitutes all of
the observed DM, Eq.~\eqref{eq:relic} must reproduce
$\Omega_{\rm DM}h^2=0.120\pm0.001$~\cite{Planck:2018vyg}. Once the Standard
Model thermal history is fixed, the abundance depends, apart from the
effective couplings, on the DM mass $m_\chi$ and the reheating temperature
$T_R$.

\medskip
Before proceeding with the numerical analysis, it is useful to isolate the
dependence of the relic abundance on the reheating temperature. For this
purpose, we neglect the temperature dependence of $g_*$ and $g_{*s}$.
We focus here on vector bilinears, \textit{i.e.} $X=V,A,L,R$, although analogous results are expected
for the other Lorentz structures. In this case, the total decay rate and the
spin-averaged annihilation cross section are
\begin{equation}\label{eq:Gamma&XS}
\begin{aligned}
\Gamma_{\mu\to e\chi\bar\chi}
&\simeq
\frac{m_\mu^5}
     {384\pi^3\left(\Lambda^{\mu e}_{X}\right)^4}
&&\Longrightarrow&
{\cal C}_{1\to3}(T)
&\simeq
\frac{m_\mu^8}
     {192\pi^5\left(\Lambda^{\mu e}_{X}\right)^4}
\frac{K_1(x)}{x}\,,
\\[2mm]
\sigma_{\mu\bar e\to\chi\bar\chi}(y)
&\simeq
\frac{m_\mu^2}
     {24\pi\left(\Lambda^{\mu e}_{X}\right)^4}
\left(2y^2+1\right)
&&\Longrightarrow&
{\cal C}_{2\to2}(T)
&\simeq
\frac{m_\mu^8}
     {48\pi^5\left(\Lambda^{\mu e}_{X}\right)^4}
\frac{K_\sigma(x)}{x}\,.
\end{aligned}
\end{equation}
where
$K_\sigma(x)\equiv\int_1^{\infty}{\rm d}y\,
\left(y^2-1\right)^2\left(2y^2+1\right)
K_1\left(xy\right)$.
In Eq.~\eqref{eq:Gamma&XS}, the approximate expressions hold in the limit
$m_\chi\ll m_\mu$. The decay and annihilation contributions to the relic
abundance, denoted by $\left.\Omega_\chi h^2\right|_{1\to3}$ and
$\left.\Omega_\chi h^2\right|_{2\to2}$, respectively, can then be expressed as
\begin{equation}\label{eq:relic-both}
\begin{split}
&\left.\Omega_\chi h^2\right|_{1\to3}
=
\frac{1}{128\pi^4}\,
\frac{2.8M_{\rm Pl}\mathcal S_0}
{g_{*s}\sqrt{g_*}\,\rho_c^0/h^2}
\frac{m_\mu^3m_\chi}
{\left(\Lambda_X^{\mu e}\right)^4}
\, {\cal F}_{1\to3}(x_R)\,,
\\
&\left.\Omega_\chi h^2\right|_{2\to2}
=
\frac{1}{32\pi^4}\,
\frac{2.8M_{\rm Pl}\mathcal S_0}
{g_{*s}\sqrt{g_*}\,\rho_c^0/h^2}
\frac{m_\mu^3m_\chi}
{\left(\Lambda_X^{\mu e}\right)^4}
\, {\cal F}_{2\to2}(x_R)\,,
\end{split}
\end{equation}
where
${\cal F}_{1\to3}(x_R)\equiv
2/(3\pi)\int_{x_R}^{\infty}{\rm d}x\,x^3K_1(x)$
and
${\cal F}_{2\to2}(x_R)\equiv
2/(3\pi)\int_{x_R}^{\infty}{\rm d}x\,x^3K_\sigma(x)$.

\smallskip
For $T_R\gg m_\mu>m_\chi$, one has $x_R\to0$ and therefore ${\cal F}_{1\to3}(x_R)\to1$. By contrast, the annihilation contribution is dominated by the large-$y$ region of $K_\sigma(x)$ and behaves as ${\cal F}_{2\to2}(x_R)\simeq512/(3\pi x_R^3)$. Using $\mathcal S_0\simeq2891.2\,{\rm cm}^{-3}$ and $\rho_c^0/h^2=1.054\times10^{-5}\,{\rm GeV\,cm}^{-3}$ for the present entropy and critical densities, respectively, and neglecting the difference between $g_*$ and $g_{*s}$, one obtains
\begin{equation}\label{eq:analxR0}
\begin{split}
\left.\Omega_\chi h^2\right|_{1\to3}
&\simeq
0.12
\left(\frac{18}{g_*}\right)^{3/2}
\left(\frac{m_\chi}{\MeV}\right)
\left(\frac{17.6\,\TeV}{\Lambda_X^{\mu e}}\right)^4
\,,
\\
\left.\Omega_\chi h^2\right|_{2\to2}
&\simeq
0.12
\left(\frac{18}{g_*}\right)^{3/2}
\left(\frac{m_\chi}{\MeV}\right)
\left(\frac{67.5\,\TeV}{\Lambda_X^{\mu e}}\right)^4\left(\frac{T_R}{m_\mu}\right)^3
\,.
\end{split}
\end{equation}
showing explicitly that DM production from muon decays is IR dominated and becomes independent of the reheating temperature for $T_R\gg m_\mu$, whereas the annihilation contribution is UV dominated and scales as $T_R^3$.

In the opposite limit, $m_\chi<T_R\ll m_\mu$, one has $x_R\to\infty$. Expanding the modified Bessel function for large arguments, one obtains ${\cal F}_{1\to3}(x_R)\simeq 1/3\sqrt{2/\pi}\,x_R^{5/2}e^{-x_R}$, which displays the Boltzmann suppression associated with the small thermal abundance of muons. The annihilation integral is instead dominated by the region $y\simeq1$, yielding ${\cal F}_{2\to2}(x_R)\simeq 24\,{\cal F}_{1\to3}(x_R)/x_R^3$. Thus, in addition to the common Boltzmann suppression, the annihilation contribution is suppressed by three further powers of $T_R/m_\mu$ relative to the decay contribution. Neglecting the difference between $g_*$ and $g_{*s}$, one finds
\begin{equation}\label{eq:analxRinf}
\begin{split}
\left.\Omega_\chi h^2\right|_{1\to3}
&\simeq
0.12
\left(\frac{11}{g_*}\right)^{3/2}
\left(\frac{m_\chi}{\MeV}\right)
\left(\frac{4.73\,\TeV}{\Lambda_X^{\mu e}}\right)^4
\left(\frac{10 \,\MeV}{T_R}\right)^{5/2}
e^{-\frac{10 \MeV}{T_R}}
\,,
\\
\left.\Omega_\chi h^2\right|_{2\to2}
&\simeq
0.12
\left(\frac{11}{g_*}\right)^{3/2}
\left(\frac{m_\chi}{\MeV}\right)
\left(\frac{2.53\,\TeV}{\Lambda_X^{\mu e}}\right)^4
\left(\frac{T_R}{10 \,\MeV}\right)^{1/2}
e^{-\frac{10 \MeV}{T_R}}
\, .
\end{split}
\end{equation}

\begin{figure}
    \centering
    \includegraphics[width=0.49\linewidth]{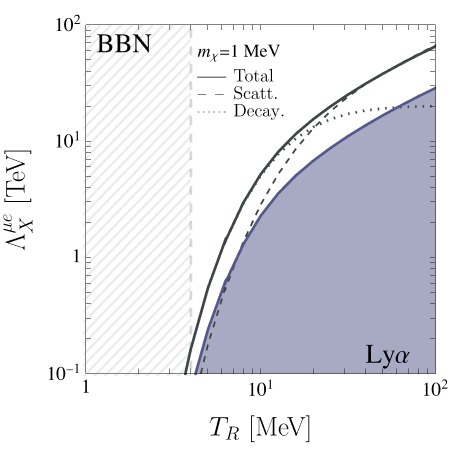}
        \includegraphics[width=0.49\linewidth]{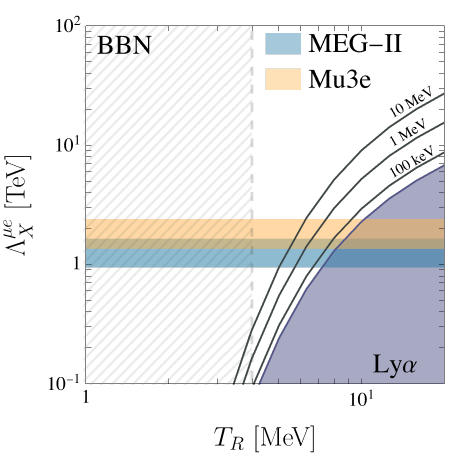}
        \caption{Parameter regions in the $(T_R,\Lambda_X^{\mu e})$ plane in which the LFV operator reproduces the observed DM relic abundance through freeze-in production. (Left panel) Contributions from muon decay (dotted), annihilation (dashed), and their sum (solid), for $m_\chi=1\,\MeV$. (Right panel) Total freeze-in contribution for $m_\chi=100\,\keV$, $1\,\MeV$, and $10\,\MeV$. The orange and blue shaded regions indicate the approximate $95\%$ CL sensitivities of Mu3e and MEG~II, respectively, for the $X=V,A$ scenarios. In both panels, the hatched gray region is excluded by the BBN bound $T_R\gtrsim4\,\MeV$, while the purple shaded region is excluded by the indicative Ly-$\alpha$ lower bound $m_\chi\gtrsim37\,\keV$, translated into the $(T_R,\Lambda_X^{\mu e})$ plane.}
    \label{fig:totalTRLambda}
\end{figure}

\medskip
The analytical solutions derived above, obtained by neglecting the
$x$ dependence of $g_*$ and $g_{*s}$, and for $m_\chi\ll m_\mu$, provide an excellent approximation to the full numerical solution of the Boltzmann equation in Eq.~\eqref{eq:boltzmann}, with only mild deviations over the parameter range of interest. Nevertheless, in the numerical analysis below we retain both the full dependence on the DM mass and the complete temperature dependence $g_*$ and $g_{*s}$. As anticipated from the analytical estimates, imposing that freeze-in production
reproduces the observed DM abundance selects a definite relation among $T_R$, $\Lambda_X^{\mu e}$, and $m_\chi$. 

\smallskip
Figure~\ref{fig:totalTRLambda} explores the $(T_R,\Lambda_X^{\mu e})$ plane for all the vector bilinear structures $X=V,A,L,R$, fixing the DM mass to representative values. 

The left panel illustrates the interplay between the $1\to3$ decay and the $2\to2$ annihilation processes for $m_\chi=1\,\MeV$. The solid black curve shows the value of
$\Lambda_X^{\mu e}$ required to reproduce the observed relic abundance when both production mechanisms are included, while the dotted and dashed black curves show the corresponding results obtained from decay and annihilation alone, respectively. Since the two contributions add at the level of the abundance, the effective scale obtained from their sum lies above those associated with either individual channel and approaches the curve
corresponding to the dominant production mechanism. Two different regimes can be identified:
\begin{itemize}
\item For $T_R> m_\mu$, the annihilation contribution dominates because of its UV sensitivity. As shown in Eq.~\eqref{eq:analxR0}, the corresponding relic abundance scales as $T_R^3$ and, therefore, the value of $\Lambda_X^{\mu e}$ required to reproduce the observed abundance scales as $T_R^{3/4}$. By contrast, the decay contribution saturates once the thermal bath reaches temperatures above the muon mass, and the corresponding
effective scale becomes approximately independent of $T_R$.

\item For $T_R<m_\mu$, the thermal muon abundance becomes Boltzmann
suppressed and both production mechanisms rapidly lose efficiency. The annihilation contribution is additionally suppressed relative to the decay one, so DM production becomes increasingly dominated by $\mu\to e\chi\bar\chi$. Accordingly, the effective scale required to reproduce the observed abundance decreases with $T_R$ and follows the asymptotic behavior
$\Lambda_X^{\mu e}\propto x_R^{5/8}e^{-x_R/4}$.
\end{itemize}
The transition between the annihilation- and decay-dominated regimes occurs at reheating temperatures of order a few tens of MeV.

The right panel of Fig.~\ref{fig:totalTRLambda} shows the combined freeze-in solutions for $m_\chi=0.1\,\MeV$, $1\,\MeV$, and $10\,\MeV$. Away from the kinematic endpoint, the yield depends only mildly on $m_\chi$, while the relic density scales as
$\Omega_\chi h^2\propto m_\chi/(\Lambda_X^{\mu e})^4$, as follows from Eq.~\eqref{eq:relic-both}. The effective scale required to reproduce the observed abundance therefore scales approximately as
$\Lambda_X^{\mu e}\propto m_\chi^{1/4}$. The three curves consequently have a similar dependence on $T_R$ and shift towards larger values of $\Lambda_X^{\mu e}$ as $m_\chi$ increases. 

The orange and blue regions show the projected $95\%$ CL sensitivities of Mu3e and MEG~II, respectively, for $X=V,A$. As already discussed in relation to Fig.~\ref{fig:Bound_inv_search}, for DM masses below $10~\mathrm{MeV}$ the constraints are broadly comparable across the different scenarios, with the exception of the $X=L$ case. In this mass range, they also provide a useful comparison between the sensitivity of LFV experiments and cosmological bounds. Their overlap with the freeze-in curves identifies the parameter space in which the same LFV interaction can generate the observed DM abundance and be tested in rare muon decays. This occurs mainly at low reheating temperatures, where annihilation is suppressed and production is dominated by $\mu\to e\chi\bar\chi$. 

The relevant region remains compatible with the BBN bound $T_R\gtrsim4\,\MeV$~\cite{Hannestad:2004px}, shown by the
hatched gray region. Lighter DM masses correspond to smaller effective scales and therefore allow an overlap with the experimental sensitivities over a broader range of $T_R$.

\medskip
It is also useful to present the results in the $(m_\chi,\Lambda_X^{\mu e})$ plane, as shown in Fig.~\ref{fig:totalmchiLambda}. The left and right panels display the projected $95\%$ CL sensitivities of MEG~II and Mu3e, respectively, based on the analyses presented in Figs.~\ref{fig:Bound_inv_search} and~\ref{fig:RMDbound} and combining the positron-energy and $\cos\theta_e$ distributions.

For MEG~II, the purple, blue, and orange bands show the projected bounds from $\mu\to e\chi\bar\chi$ for the $X=V,A$, $X=R$, and $X=L$ scenarios, respectively. Their widths correspond to varying the number of muon decays between $10^5$ and $10^7$, while fixing the systematic uncertainty at the per-mille level. The solid curves indicate the bounds from the radiative channel $\mu\to e\chi\bar\chi\gamma$, obtained for $10^5$ muon decays and a systematic uncertainty of $4\%$. For Mu3e, the number of muon decays is fixed to $10^{10}$, while the systematic uncertainty is varied between $10^{-3}$ and $10^{-2}$. The black dashed curves show the values of $\Lambda_X^{\mu e}$ required to reproduce the observed relic abundance for $T_R=6,\,8,$ and $10\,\MeV$. In this regime, freeze-in production is dominated by muon decays and the thermal curves approximately follow $\Lambda_X^{\mu e}\propto m_\chi^{1/4}$. Lower reheating temperatures lead to a stronger Boltzmann suppression of the thermal muon abundance and therefore require smaller effective scales, with the non-linear dependence on $T_R$ described by Eq.~\eqref{eq:analxRinf}. The overlap between the thermal curves and the projected sensitivities directly identifies the range of DM masses for which freeze-in production can be tested by MEG~II and Mu3e.

\begin{figure}
    \centering
    \includegraphics[width=0.49\linewidth]{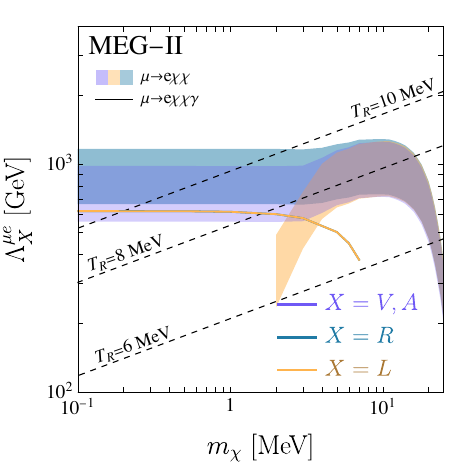}
        \includegraphics[width=0.49\linewidth]{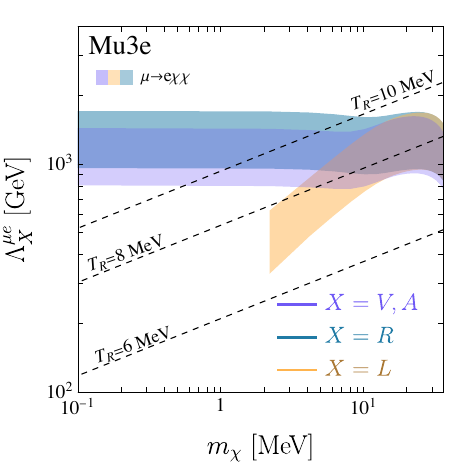}
        \caption{95\% CL constraints on the effective NP scale $\Lambda^{\mu e}_X$ from MEG-II (left panel) and Mu3e (right panel), corresponding to the analyses shown in Figs.~\ref{fig:Bound_inv_search} and~\ref{fig:RMDbound}. For MEG-II, the purple, blue, and orange bands show the constraints from $\mu\to e\chi\chi$ analysis for the $X=V,A$, $X=R$, and $X=L$ scenarios, respectively. The bands are obtained by varying the number of muons between $10^5$ and $10^7$, while fixing the systematic uncertainty at the per-mille level. The solid curves show instead the constraints from the radiative decay $\mu\to e\chi\chi\gamma$, obtained for $10^5$ muons and a systematic uncertainty of $4\%$. For Mu3e, the number of muons is fixed to $10^{10}$, while the systematic uncertainty is varied between $10^{-3}$ and $10^{-2}$. In both experiments, the analysis combines binning in the positron energy and in $\cos\theta_e$. The black dashed curves indicate the thermal benchmarks corresponding to reheating temperatures of $6$, $8$, and $10~\mathrm{MeV}$ and are independent of the interaction scenario considered.}
    \label{fig:totalmchiLambda}
\end{figure}

\medskip
The thermal warm Dark Matter (WDM) bound can be recast for non-thermal freeze-in scenarios by comparing the corresponding DM momentum distributions
\cite{DEramo:2020gpr,Ballesteros:2020adh,Decant:2021mhj,DEramo:2025jsb}. The production of a $\chi\bar\chi$ pair through the three-body decay considered here has not been studied explicitly. We therefore adopt the indicative bound $m_\chi\gtrsim37\,\keV$ derived in Ref.~\cite{DEramo:2025jsb} for two-body decays producing a single DM particle. A constraint of similar size was found for scattering production in the range $4\,\MeV\lesssim T_R\lesssim100\,\MeV$. Although we do not expect a substantial modification in our setup, a precise limit would require a
dedicated calculation of the momentum distribution generated by three-body decays and by the combined decay and annihilation contributions.

For fixed $T_R$, the lower bound on $m_\chi$ translates into a lower bound on the value of $\Lambda_X^{\mu e}$ required to reproduce the observed relic abundance. The corresponding region is shown in purple in Figs.~\ref{fig:totalTRLambda} and should be regarded as an indicative estimate of the parameter space disfavored by structure-formation constraints.

\bigskip
To conclude this section, let us briefly discuss the possible impact of flavor-diagonal operators. Their contribution simply adds to the relic abundance obtained from the LFV interaction,
\begin{equation}
\Omega_\chi h^2
=
\left.\Omega_\chi h^2\right|_{\mu\to e\chi\bar\chi}
+
\left.\Omega_\chi h^2\right|_{\mu\bar e\to\chi\bar\chi}
+
\sum_{\ell=e,\mu}
\left.\Omega_\chi h^2\right|_{\ell\bar\ell\to\chi\bar\chi}\,.
\end{equation}
For a flavor-diagonal operator, the collision term is a factor of two smaller than for the LFV annihilation process considered above. Indeed, the LFV contribution includes the two distinct charge-conjugate channels $\mu^-e^+\to\chi\bar\chi$ and $\mu^+e^-\to\chi\bar\chi$, whereas $\ell^-\ell^+\to\chi\bar\chi$ represents a single initial-state channel. The corresponding contribution to the relic abundance is therefore
\begin{equation}
\left.\Omega_\chi h^2\right|_{\ell\bar\ell\to\chi\bar\chi}
=
\frac{1}{64\pi^4}
\frac{2.8M_{\rm Pl}{\cal S}_0h^2}
{g_{*s}\sqrt{g_*}\rho_c^0}
\frac{m_\ell^3m_\chi}
{\left(\Lambda_X^{\ell\ell}\right)^4}
\,{\cal F}_{2\to2}^{\ell\ell}(x_R^\ell)\,,
\end{equation}
where ${\cal F}_{2\to2}^{\ell\ell}(x_R^\ell)
\equiv
2/(3\pi)
\int_{x_R^\ell}^{\infty}{\rm d}x\,
x^3K_\sigma^{\ell\ell}(x)$ and
$K_\sigma^{\ell\ell}(x)
\equiv
\int_2^\infty {\rm d}y\,
y^3\left(y^2+2\right)\sqrt{y^2-4}\,
K_1(xy)$. Here, the dimensionless variables are defined as in the LFV case, with the important replacement $m_\mu\to m_\ell$. Neglecting $m_\chi$, $K_\sigma^{\ell\ell}$ is the same dimensionless function for electrons and muons, with the dependence on the lepton species entering only through $x=m_\ell/T$ and $x_R^\ell=m_\ell/T_R$. Its expression differs from the LFV kernel because the equal-mass initial state is governed by the Källén function $\lambda(s,m_\ell^2,m_\ell^2)=s(s-4m_\ell^2)$, rather than by $\lambda(s,m_\mu^2,0)=(s-m_\mu^2)^2$.

\medskip
For the muon flavor-diagonal operator, the annihilation process $\mu^+\mu^-\to\chi\bar\chi$ is even more strongly suppressed than the LFV decay in the region $T_R<m_\mu$. Indeed, it requires two thermally populated muons and therefore scales as ${\cal F}_{2\to2}^{\mu\mu}\propto e^{-2m_\mu/T_R}$, whereas the decay contribution contains only the single Boltzmann suppression $e^{-m_\mu/T_R}$.

\medskip
The situation is qualitatively different for the electron operator. In the reheating-temperature range of interest, one typically has $T_R\gg m_e$, so electrons remain relativistic and their annihilation contribution is not Boltzmann suppressed. In this limit, ${\cal F}_{2\to2}^{ee}\propto(T_R/m_e)^3$, and the factor $m_e^3$ in the prefactor cancels, leaving an abundance proportional to $T_R^3$. Therefore, if $\Lambda_X^{ee}$ is comparable to $\Lambda_X^{\mu e}$, the process $e^+e^-\to\chi\bar\chi$ can significantly enhance, or even dominate, the total freeze-in abundance. This is the same $T_R^3$ dependence found for the LFV annihilation contribution in Eq.~\eqref{eq:analxR0}, but for electrons the relativistic approximation applies throughout the reheating-temperature range relevant for LFV experiments. One obtains
\begin{equation}\label{eq:analxR0ee}
\left.\Omega_\chi h^2\right|_{e^+e^-\to\chi\bar\chi}
\simeq
0.12
\left(\frac{11}{g_*}\right)^{3/2}
\left(\frac{m_\chi}{\MeV}\right)
\left(\frac{11.7\,\TeV}{\Lambda_X^{ee}}\right)^4
\left(\frac{T_R}{10\,{\rm MeV}}\right)^3
\,.
\end{equation}
This expression explicitly shows that, for comparable effective scales, the freeze-in abundance is driven by the flavor-diagonal electron interaction. In particular, throughout the low-reheating-temperature region relevant for LFV searches, electron annihilation remains efficient, whereas production from muon decays and muon-initiated annihilations is increasingly suppressed as $T_R$ decreases below $m_\mu$. The connection between the observed relic abundance and the LFV operator therefore remains direct only if the electron coupling is sufficiently suppressed, corresponding approximately to $\Lambda_X^{ee}\gg\Lambda_X^{\mu e}$. A possible flavor structure realizing such a hierarchy is discussed in Appendix~\ref{App:FSE}. 

\section{Conclusions}\label{sec:conclusions}
Searches for charged-lepton flavor-violating processes are among the most
powerful probes of new physics beyond the Standard Model and constitute a
major component of the current high-intensity experimental program. In this
work, we have investigated the prospects of MEG~II and Mu3e for detecting
LFV muon decays into a positron and a pair of stable neutral fermions that
may constitute dark matter. Within an effective-field-theory framework, we
parametrized these interactions in terms of dimension-six four-fermion
operators and considered all possible Lorentz structures. Particular
attention was devoted to vector interactions, which arise naturally from
the exchange of a heavy neutral mediator in weakly coupled ultraviolet
completions.

We characterized the signal kinematics and performed a shape analysis of the
Michel spectrum under realistic experimental assumptions. Our results show
that the decay $\mu\to e\chi\bar\chi$ can be distinguished from the SM
background, allowing MEG~II and Mu3e to probe effective new-physics scales
of ${\cal O}({\rm TeV})$ for DM masses above approximately
$1~{\rm MeV}$. For vector interactions, the sensitivity to the chiral
structure is driven primarily by the positron angular distribution, whereas
the energy spectrum is largely insensitive to the distinction among pure
vector, pure axial-vector, left-handed, and right-handed scenarios. As a
consequence, the bounds approach approximately constant values at small
$m_\chi$ and are determined predominantly by angular information. Pure
vector and pure axial-vector interactions lead to approximately isotropic
distributions that remain clearly distinguishable from the SM Michel
spectrum. A purely right-handed interaction instead produces an angular
slope opposite to the SM one and therefore provides the strongest
discrimination power. By contrast, in the purely left-handed scenario both
the energy and angular distributions approach the SM predictions as
$m_\chi\to0$, leading to a progressive loss of sensitivity in the
non-radiative channel.

For DM masses below approximately $1~{\rm MeV}$, the signal distortion is
concentrated too close to the Michel endpoint to yield a robust constraint,
since this region is typically assumed to be signal-free and is used for
experimental calibration. We have shown that this limitation can be
overcome at MEG~II by exploiting the radiative decay
$\mu\to e\chi\bar\chi\gamma$. The additional photon and the enlarged set of
kinematic observables restore sensitivity down to the massless-DM limit,
yielding bounds that are comparable to, and in some regions stronger than,
those obtained from the non-radiative channel. The two strategies are
therefore highly complementary: the non-radiative search efficiently probes
DM masses outside the calibration-dominated region, whereas the radiative
channel provides access to very light DM and removes the degeneracy affecting
the purely left-handed scenario.

Remarkably, both MEG~II and Mu3e can probe regions of parameter space in
which the observed DM abundance is generated through freeze-in from muon
decays induced by the same LFV operators. This cosmological connection is
realized for reheating temperatures below approximately $10~{\rm MeV}$.
Taken together, these results demonstrate that precision searches for rare
muon decays provide a sensitive and complementary avenue for testing
lepton flavor-violating interactions with a dark sector, while simultaneously
probing the particle physics origin of the DM relic abundance.

\section{Acknowledgments}

We thank Alessandro Baldini, Fabrizio Cei, Alessandro Fiorentino, Luca Galli, Jacobo Lopez Pavon, Laura Molina Bueno and Diego Redigolo for useful discussions. The work of RZ is partially supported by project B3a of the DFG-funded Collaborative Research Center TRR257, “Particle Physics Phenomenology after the
Higgs Discovery" and has received support from the European Union’s Horizon 2020 research
and innovation programme under the Marie Sk{\l}odowska-Curie grant agreement No 860881-HIDDeN.
\appendix

\section{Structure of tensor interaction}\label{eq:tensor_coeff}

Here we show that the four fermion interactions involving two tensor current only admit three independent coupling structures, as opposed to the scalar and vector ones which depend on four independent parameters. Starting from the identity
\be
\gamma^5 \sigma^{\mu\nu} =  \sigma^{\mu\nu} \gamma^5 = \frac{i}{2}\varepsilon^{\mu\nu\alpha\beta} \sigma_{\alpha\beta} \ ,
\ee
then one has
\be
\begin{split}
& (\bar\psi_i \sigma^{\mu\nu} \gamma^5 \psi_j )(\bar\psi_k \sigma_{\mu\nu} \gamma^5 \psi_l ) 
= \left(\frac{i}{2}\right)^2  \varepsilon^{\mu\nu\alpha\beta}\varepsilon_{\mu\nu\rho\sigma}
(\bar\psi_i \sigma_{\alpha\beta}  \psi_j )(\bar\psi_k \sigma^{\rho\sigma}  \psi_l )  =  \\
& = +\frac{1}{2} (\delta^\alpha_\rho \delta^\beta_\sigma - \delta^\alpha_\sigma \delta^\beta_\rho) (\bar\psi_i \sigma_{\alpha\beta}  \psi_j )(\bar\psi_k \sigma^{\rho\sigma}  \psi_l ) = \\
& = + \frac{1}{2}\left(
(\bar\psi_i \sigma_{\alpha\beta}  \psi_j )(\bar\psi_k \sigma^{\alpha\beta}  \psi_l )
- (\bar\psi_i \sigma_{\alpha\beta}  \psi_j )(\bar\psi_k \sigma^{\beta\alpha}  \psi_l ) 
 \right) = 
  \, (\bar\psi_i \sigma_{\alpha\beta}  \psi_j )(\bar\psi_k \sigma^{\alpha\beta}  \psi_l ) \ ,
\end{split} 
\ee
where $i,j,k,l$ are flavor indices labelling different types of fermions and where we have used the identity
\be
\varepsilon^{\mu\nu\alpha\beta}\varepsilon_{\mu\nu\rho\sigma} = -2 (\delta^\alpha_\rho \delta^\beta_\sigma - \delta^\alpha_\sigma \delta^\beta_\rho) \ .
\ee
Analogously one has
\be
 (\bar\psi_i \sigma^{\mu\nu}  \psi_j )(\bar\psi_k \sigma_{\mu\nu} \gamma^5 \psi_l ) = 
\frac{i}{2}\epsilon_{\mu\nu\alpha\beta} (\bar\psi_i \sigma^{\mu\nu}  \psi_j ) (\bar\psi_k \sigma^{\alpha\beta}  \psi_l ) \\
=  (\bar\psi_i \sigma_{\alpha\beta} \gamma^5  \psi_j )(\bar\psi_k \sigma^{\alpha\beta}  \psi_l ) \ .
\ee
One has then the freedom to set to zero the pseudotensor coupling of either the SM lepton current or the DM current without loss of generality.

\section{Differential $\mu-$decay rates}\label{app:mu-decay}
In this Appendix, we present the differential decay rates
${\rm d}^2\Gamma/({\rm d}E_e\,{\rm d}\cos\theta_e)$ for a polarized
$\mu^+$ with polarization degree ${\cal P}$, together with the corresponding
energy spectra ${\rm d}\Gamma/{\rm d}E_e$ and total decay rates $\Gamma$.
Results are given for the SM contribution and for the three NP interactions
defined in Eq.~\eqref{eq:lag}, in the limit of a vanishing electron mass.
 
\subsection*{SM interaction}
For the SM decay $\mu^+\to e^+\nu_e\bar\nu_\mu$, we introduce the
dimensionless positron energy
\be
z_e\equiv\frac{2E_e}{m_\mu}\,.
\ee
Neglecting the neutrino masses, the kinematically allowed ranges are
$0\leq z_e\lesssim  1$ and $\cos\theta_e\in[-1,1]$. The double-differential decay rate
can be expressed in terms of the unpolarized positron-energy spectrum as
\be\label{eq:SMdoublediffrate}
\frac{{\rm d}^2\Gamma}
{{\rm d}z_e\,{\rm d}\cos\theta_e}
=
\frac{1}{2}
\frac{{\rm d}\Gamma}{{\rm d}z_e}
\left[
1+
{\cal P}\cos\theta_e\,
\frac{2z_e-1}{3-2z_e}
\right]\,.
\ee
As is apparent, the polarization-dependent term in square brackets vanishes
upon integration over $\cos\theta_e$. The unpolarized positron-energy
spectrum in Eq.~\eqref{eq:SMdoublediffrate} is
\be
\frac{{\rm d}\Gamma}{{\rm d}z_e}
=
\frac{G_F^2m_\mu^5}{96\pi^3}\,
z_e^2\left(3-2z_e\right)\,.
\ee
The total decay rate is therefore
\be
\Gamma
=
\frac{G_F^2m_\mu^5}{96\pi^3}
\int_0^1{\rm d}z_e\,
z_e^2\left(3-2z_e\right)
=
\frac{G_F^2m_\mu^5}{192\pi^3}\,.
\ee

\subsection*{LFV interactions}

To express the decay distributions for the LFV process
$\mu^+\to e^+\chi\bar\chi$ in a compact form, we retain the definition of
$z_e$ introduced above and define the dimensionless mass ratio
\be
r_\chi\equiv\frac{m_\chi}{m_\mu}\,.
\ee
The kinematically allowed ranges are
$0\leq z_e\leq1-4r_\chi^2$ and $\cos\theta_e\in[-1,1]$. We also introduce
the following combinations of Wilson coefficients for the scalar, vector,
and tensor interactions:
\be
\begin{alignedat}{3}
\Sigma_{\rm scalar}^{\ell,\chi}
&\equiv |c_S^{\ell,\chi}|^2+|c_P^{\ell,\chi}|^2\,,
&\qquad
\Sigma_{\rm vector}^{\ell,\chi}
&\equiv |c_V^{\ell,\chi}|^2+|c_A^{\ell,\chi}|^2\,,
&\qquad
\Sigma_{\rm tensor}^{\ell}
&\equiv |c_T^\ell|^2+|c_{T'}^\ell|^2\,,
\\[1mm]
\Delta_{\rm scalar}^{\ell,\chi}
&\equiv |c_S^{\ell,\chi}|^2-|c_P^{\ell,\chi}|^2\,,
&
\Delta_{\rm vector}^{\ell,\chi}
&\equiv |c_V^{\ell,\chi}|^2-|c_A^{\ell,\chi}|^2\,,
&
\Delta_{\rm tensor}^{\ell}
&\equiv |c_T^\ell|^2-|c_{T'}^\ell|^2\,,
\\[1mm]
\Pi_{\rm scalar}^{\ell}
&\equiv {\rm Re}\!\left[c_S^\ell c_P^{\ell,*}\right]\,,
&
\Pi_{\rm vector}^{\ell}
&\equiv {\rm Re}\!\left[c_V^\ell c_A^{\ell,*}\right]\,,
&
\Pi_{\rm tensor}^{\ell}
&\equiv {\rm Re}\!\left[c_T^\ell c_{T'}^{\ell,*}\right]\,.
\end{alignedat}
\ee
With these definitions, the double-differential decay rate can be written in
terms of the unpolarized positron-energy spectrum as
\be\label{eq:LFVdoublediffrate}
\frac{{\rm d}^2\Gamma_{\cal I}}
{{\rm d}z_e\,{\rm d}\cos\theta_e}
=
\frac{1}{2}
\frac{{\rm d}\Gamma_{\cal I}}{{\rm d}z_e}
\left[
1+
2{\cal P}\cos\theta_e\,
\frac{\Pi_{\cal I}^\ell}
{\Sigma_{\cal I}^\ell}
\frac{
\mathcal W_{\cal I}(z_e,r_\chi)
}{
\mathcal U_{\cal I}(z_e,r_\chi)
}
\right]\,,
\ee
where ${\cal I}=\{{\rm scalar},{\rm vector},{\rm tensor}\}$. As in the SM
decay, the polarization-dependent term in square brackets vanishes upon
integration over $\cos\theta_e$. The unpolarized positron-energy spectrum
appearing in Eq.~\eqref{eq:LFVdoublediffrate} is
\be\label{eq:LFVdiffrate}
\frac{{\rm d}\Gamma_{\cal I}}{{\rm d}z_e}
=
\frac{\Sigma_{\cal I}^\ell m_\mu^5}
{96\pi^3\Lambda^4}\,
z_e^2
\sqrt{1-\frac{4r_\chi^2}{1-z_e}}\,
\mathcal U_{\cal I}(z_e,r_\chi)\,.
\ee
The corresponding total decay rate is therefore
\be\label{eq:LFVtotalrate}
\Gamma_{\cal I}
=
\frac{\Sigma_{\cal I}^\ell m_\mu^5}
{96\pi^3\Lambda^4}
\int_0^{1-4r_\chi^2}
{\rm d}z_e\,
z_e^2
\sqrt{1-\frac{4r_\chi^2}{1-z_e}}\,
\mathcal U_{\cal I}(z_e,r_\chi)\,.
\ee

\medskip

The dependence on the Lorentz structure is entirely encoded in the
dimensionless kinematic functions. For the scalar interaction, they are
\be\label{eq:UWscalar}
\mathcal U_{\rm scalar}(z_e,r_\chi)
=
\mathcal W_{\rm scalar}(z_e,r_\chi)
\equiv
\frac{3}{4}(1-z_e)\,
\Sigma_{\rm scalar}^\chi
-
\frac{3}{2}r_\chi^2
\left(
\Sigma_{\rm scalar}^\chi+\Delta_{\rm scalar}^\chi
\right)\,.
\ee
For the vector interaction, one obtains
\be\label{eq:UWvector}
\begin{split}
\mathcal U_{\rm vector}(z_e,r_\chi)
&\equiv
3r_\chi^2\Delta_{\rm vector}^\chi
+
\frac{
2z_e^2+\left(2r_\chi^2-5\right)z_e+3
}{
2(1-z_e)
}\,
\Sigma_{\rm vector}^\chi\,,
\\
\mathcal W_{\rm vector}(z_e,r_\chi)
&\equiv
3r_\chi^2\Delta_{\rm vector}^\chi
+
\frac{
2z_e^2+\left(2r_\chi^2-3\right)z_e
+1-4r_\chi^2
}{
2(1-z_e)
}\,
\Sigma_{\rm vector}^\chi\,.
\end{split}
\ee
Finally, for the tensor interaction, the kinematic functions are
\be
\label{eq:UWtensor}
\begin{split}
\mathcal U_{\rm tensor}(z_e,r_\chi)
&\equiv
2|c_T^\chi|^2
\frac{
z_e^2-2\left(2+r_\chi^2\right)z_e
+3\left(1+2r_\chi^2\right)
}{
1-z_e
}\,,
\\
\mathcal W_{\rm tensor}(z_e,r_\chi)
&\equiv
2|c_T^\chi|^2
\frac{
z_e^2-2r_\chi^2z_e-1-2r_\chi^2
}{
1-z_e
}\,.
\end{split}
\ee
Inserting these functions into Eq.~\eqref{eq:LFVtotalrate} and carrying out
the integration over $z_e$ gives, for the scalar interaction,
\be\label{eq:LFVtotal_scalar}
\begin{split}
\Gamma_{\rm scalar}
={}&
\frac{\Sigma_{\rm scalar}^{\ell}m_\mu^5}
{1536\pi^3\Lambda^4}
\Bigg\{
\sqrt{1-4r_\chi^2}
\Bigg[
\Sigma_{\rm scalar}^{\chi}
\left(
1-14r_\chi^2-2r_\chi^4-12r_\chi^6
\right)
-
\Delta_{\rm scalar}^{\chi}
\left(
8r_\chi^2+40r_\chi^4-48r_\chi^6
\right)
\Bigg]
\\
&+
48r_\chi^4
\operatorname{arctanh}
\left(
\sqrt{1-4r_\chi^2}
\right)
\Bigg[
\Sigma_{\rm scalar}^{\chi}
\left(
1-r_\chi^4
\right)
+
\Delta_{\rm scalar}^{\chi}
\left(
2-4r_\chi^2+4r_\chi^4
\right)
\Bigg]
\Bigg\}\,.
\end{split}
\ee
For the vector interaction, the total rate is
\be\label{eq:LFVtotal_vector}
\begin{split}
\Gamma_{\rm vector}
={}&
\frac{\Sigma_{\rm vector}^{\ell}m_\mu^5}
{384\pi^3\Lambda^4}
\Bigg\{
\sqrt{1-4r_\chi^2}
\Bigg[
\Sigma_{\rm vector}^{\chi}
\left(
1-14r_\chi^2-2r_\chi^4-12r_\chi^6
\right)
+
\Delta_{\rm vector}^{\chi}
\left(
4r_\chi^2+20r_\chi^4-24r_\chi^6
\right)
\Bigg]
\\
&+
48r_\chi^4
\operatorname{arctanh}
\left(
\sqrt{1-4r_\chi^2}
\right)
\Bigg[
\Sigma_{\rm vector}^{\chi}
\left(
1-r_\chi^4
\right)
+
\Delta_{\rm vector}^{\chi}
\left(
-1+2r_\chi^2-2r_\chi^4
\right)
\Bigg]
\Bigg\}\,.
\end{split}
\ee
For the tensor interaction, one finds
\be\label{eq:LFVtotal_tensor}
\begin{split}
\Gamma_{\rm tensor}
={}&
\frac{
\Sigma_{\rm tensor}^{\ell}m_\mu^5
}{
64\pi^3\Lambda^4
}
\Bigg\{
\sqrt{1-4r_\chi^2}
\,|c_T^\chi|^2\left(
1-14r_\chi^2-2r_\chi^4-12r_\chi^6
\right)
\\
&+
48r_\chi^4
\operatorname{arctanh}
\left(
\sqrt{1-4r_\chi^2}
\right) |c_T^\chi|^2 \left(
1-r_\chi^4
\right)
\Bigg\}\,.
\end{split}
\ee

In the massless-DM limit, $r_\chi\to0$, the phase-space factor approaches
unity and the endpoint of the positron-energy spectrum becomes $z_e=1$.
The kinematic functions and their relevant integrals then simplify as
follows. For the scalar interaction, one finds
\be
\begin{aligned}
&\mathcal U_{\rm scalar}(z_e,0)
=
\frac{3}{4}(1-z_e)\Sigma_{\rm scalar}^\chi
\quad \mbox{and} \quad
\frac{\mathcal W_{\rm scalar}(z_e,0)}
{\mathcal U_{\rm scalar}(z_e,0)}
=1\,,
\\[1mm]
&\int_0^1{\rm d}z_e\,z_e^2\,
\mathcal U_{\rm scalar}(z_e,0)
=
\frac{\Sigma_{\rm scalar}^\chi}{16}
\quad\Longrightarrow\quad
\Gamma_{\rm scalar}
=
\frac{
\Sigma_{\rm scalar}^\ell
\Sigma_{\rm scalar}^\chi m_\mu^5
}{
1536\pi^3\Lambda^4
}\,.
\end{aligned}
\label{eq:uscalar}
\ee
For the vector interaction,
\be
\begin{aligned}
&\mathcal U_{\rm vector}(z_e,0)
=
\frac{1}{2}(3-2z_e)\Sigma_{\rm vector}^\chi
\quad \mbox{and} \quad
\frac{\mathcal W_{\rm vector}(z_e,0)}
{\mathcal U_{\rm vector}(z_e,0)}
=
\frac{1-2z_e}{3-2z_e}\,,
\\[1mm]
&\int_0^1{\rm d}z_e\,z_e^2\,
\mathcal U_{\rm vector}(z_e,0)
=
\frac{\Sigma_{\rm vector}^\chi}{4}
\quad\Longrightarrow\quad
\Gamma_{\rm vector}
=
\frac{
\Sigma_{\rm vector}^\ell
\Sigma_{\rm vector}^\chi m_\mu^5
}{
384\pi^3\Lambda^4
}\,.
\end{aligned}
\label{eq:uvector}
\ee
Finally, for the tensor interaction,
\be\label{eq:tensorm0limit}
\begin{aligned}
&\mathcal U_{\rm tensor}(z_e,0)
=
2(3-z_e)|c_T^\chi|^2
\quad \mbox{and} \quad
\frac{\mathcal W_{\rm tensor}(z_e,0)}
{\mathcal U_{\rm tensor}(z_e,0)}
=
-\frac{1+z_e}{3-z_e}\,,
\\[1mm]
&\int_0^1{\rm d}z_e\,z_e^2\,
\mathcal U_{\rm tensor}(z_e,0)
=
\frac{3}{2}|c_T^\chi|^2
\quad\Longrightarrow\quad
\Gamma_{\rm tensor}
=
\frac{
\Sigma_{\rm tensor}^\ell
|c_T^\chi|^2m_\mu^5
}{
64\pi^3\Lambda^4
}\,.
\end{aligned}
\ee
Together with Eqs.~\eqref{eq:LFVdoublediffrate} and~\eqref{eq:LFVdiffrate}, these results make it straightforward to recover the massless-limit decay distributions given in Eq.~\eqref{eq:rates}.

\section{Results for scalar and tensor interactions}\label{app:distr_tensor}
For the scalar interaction, we consider four benchmark scenarios. The pure scalar and pure pseudoscalar cases, denoted by $X=S$ and $X=P$, are defined by $c_P^{\ell,\chi}=0$ and $c_S^{\ell,\chi}=0$, respectively. We also consider the combinations $X=S^+$ and $X=S^-$, corresponding to $c_S^{\ell,\chi}=+c_P^{\ell,\chi}$ and $c_S^{\ell,\chi}=-c_P^{\ell,\chi}$, respectively. In analogy with the change to the left- and right-handed basis adopted for the vector interaction, the associated Wilson coefficients are defined as $c_{S^+}^{\ell,\chi}/\sqrt{2}\equiv c_S^{\ell,\chi}=c_P^{\ell,\chi}$ or $c_{S^-}^{\ell,\chi}/\sqrt{2}\equiv c_S^{\ell,\chi}=-c_P^{\ell,\chi}$. As a result, the corresponding effective scales in the $S^\pm$ basis are given by $\Lambda_{S^\pm}^{\mu e}\equiv\Lambda/\sqrt{c_{S^\pm}^{\ell}c_{S^\pm}^{\chi}}$. For the tensor interaction, since $c_T^\chi$ is the only Wilson coefficient available on the DM side, we consider three benchmark scenarios. The pure tensor case, denoted by $X=T$, is defined by $c_{T'}^\ell=0$, while the combinations $X=T^+$ and $X=T^-$ correspond to $c_T^\ell=c_{T'}^\ell$ and $c_T^\ell=-c_{T'}^\ell$, respectively. As in the previous cases, the associated Wilson coefficients can be defined as $c_{T^\pm}^\ell/\sqrt{2}\equiv c_T^\ell=\pm c_{T'}^\ell$. The corresponding effective scales in the $T^\pm$ basis are then given by $\Lambda_{T^\pm}^{\mu e}\equiv\Lambda/\sqrt{c_{T^\pm}^\ell c_T^\chi}$.

\medskip
In Fig.~\ref{fig:tensor_distr}, we show the differential decay rates as
functions of the positron energy (first column) and $\cos\theta_e$ (second
column) for the scalar (first row) and tensor (second row) interactions.

\smallskip
For scalar interactions, the positron-energy distributions become independent
of the benchmark scenario in the limit $m_\chi\to0$. Indeed, in this limit
the dimensionless kinematic function in Eq.~\eqref{eq:UWscalar} is proportional to $\Sigma_{\rm scalar}^\chi$. Once inserted into
Eq.~\eqref{eq:LFVdiffrate}, the spectrum is therefore proportional to
$\Sigma_{\rm scalar}^\ell\Sigma_{\rm scalar}^\chi/\Lambda^4$. For the
benchmark choices adopted here, this combination always reduces to the
product of the squared coupling defining the leptonic current and its DM
counterpart, with the same overall normalization in all scenarios. Finite $m_\chi$ corrections, however, introduce sizable differences among
the spectra. This behavior can again be understood directly from
Eq.~\eqref{eq:UWscalar}. The $S^+$ and $S^-$ scenarios remain degenerate,
since $\Delta_{\rm scalar}^\chi=0$ in both cases. By contrast, the
mass-dependent contribution is proportional to $\Sigma_{\rm scalar}^\chi+\Delta_{\rm scalar}^\chi =2|c_S^\chi|^2$. It therefore vanishes for a pure pseudoscalar interaction and is maximal for a pure scalar interaction, explaining the pronounced separation between their positron-energy spectra at finite $m_\chi$.  
A distinctive feature of the scalar interaction is that the angular shape is exactly independent of the DM mass. This follows immediately from Eq.~\eqref{eq:UWscalar}, since $\mathcal W_{\rm scalar}/\mathcal U_{\rm scalar}=1$ for any value of $m_\chi$. Their shape is therefore entirely determined by $\Pi_{\rm scalar}^\ell/\Sigma_{\rm scalar}^\ell$: the $S^+$ and $S^-$ benchmarks exhibit equal and opposite angular slopes, while the pure scalar and pure pseudoscalar scenarios are isotropic, since $\Pi_{\rm scalar}^\ell=0$. As a result, none of the scalar benchmark scenarios approaches the SM Michel angular distribution in the $m_\chi \to 0$ limit, shown by the black solid line.

\begin{figure}[t!]
    \centering
      
      \includegraphics[width=0.48\textwidth]{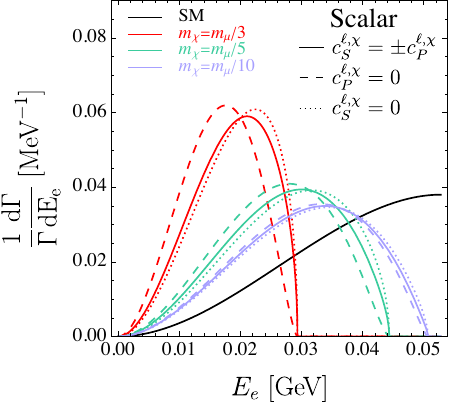}\hfill
        \includegraphics[width=0.48\textwidth]{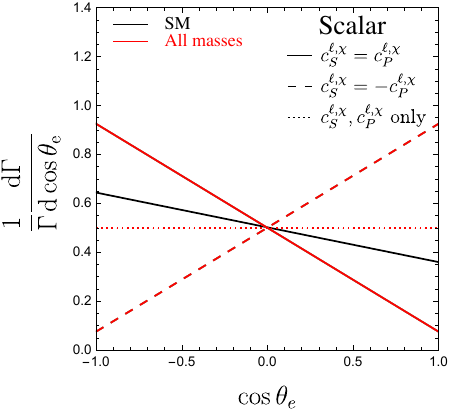} \\ \includegraphics[width=0.48\textwidth]{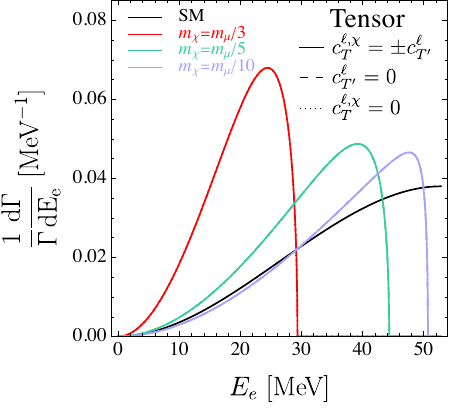}\hfill
        \includegraphics[width=0.48\textwidth]{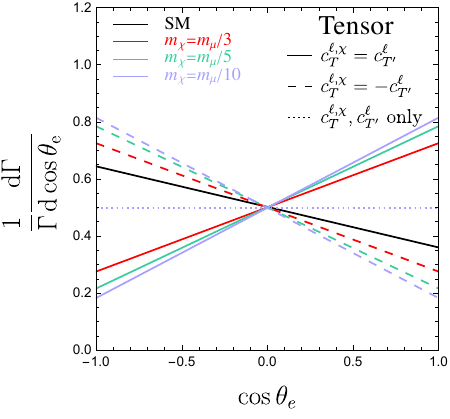} \\ 
        
    \caption{Differential rates as function of the $E_e$ and inclusive in $\cos\theta_e$ (left column) and  as function of $\cos\theta_e$ and inclusive in $E_e$ (right column) for the scalar case (top panels) and tensor case (bottom panels). In all the panels we fix the muon polarization as ${\cal P}=-0.85$. We choose $c_{T}^{\ell,\chi}=0$, $c_{ T'}^{\ell}=0$ or $c_{T}^{\ell\,\chi}=\pm c_{T'}^{\ell} $. The red, green and purple curves correspond to $m_\chi = m_\mu/3,\,  m_\mu/5$ and  $m_\mu/10$ respectively, while the black solid line represents the SM Michel spectrum.}
    \label{fig:tensor_distr}
\end{figure}

\smallskip
For tensor interactions, the kinematic function $\mathcal U_{\rm tensor}(z_e,r_\chi)$ is the same for all the tensor benchmark scenarios, and therefore the overall normalization of the spectrum is proportional to $\Sigma_{\rm tensor}^\ell |c_T^\chi|^2/\Lambda^4$. As in the scalar case, for the benchmark choices adopted here this coupling combination is normalized to the same value in all scenarios. The corresponding positron-energy distributions therefore coincide for any value of $m_\chi$. The angular distributions, by contrast, depend on the leptonic tensor structure through $\Pi_{\rm tensor}^\ell/\Sigma_{\rm tensor}^\ell$. Notice that the ratio $\mathcal W_{\rm tensor}/\mathcal U_{\rm tensor}$ is common to all tensor benchmarks; the differences among them are entirely determined by the Wilson-coefficient combination multiplying this ratio. Pure tensor and pure pseudotensor interactions lead to isotropic distributions, since $\Pi_{\rm tensor}^\ell=0$. For the mixed benchmarks, the two relative signs produce opposite angular slopes. The distribution for $c_T^\ell=-c_{T'}^\ell$ is clearly distinct from the SM Michel spectrum, whereas that for $c_T^\ell=+c_{T'}^\ell$ more closely resembles it, as can be inferred from Eq.~\eqref{eq:tensorm0limit}. Nevertheless, none of the tensor benchmarks exactly approaches the SM Michel distribution in the massless-DM limit.

\begin{figure}[t]
    \centering
    \includegraphics[width=0.49\linewidth]{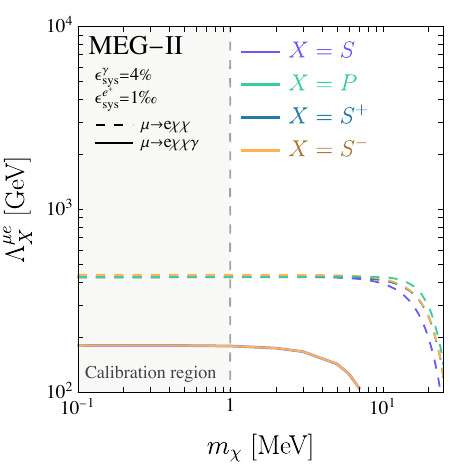}
    \hfill
    \includegraphics[width=0.49\linewidth]{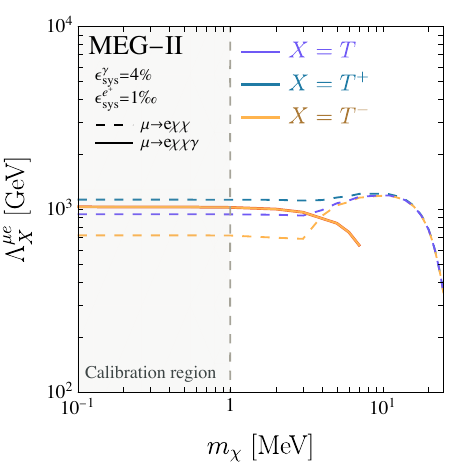}

    \caption{$95\%$ CL constraints on the effective scale $\Lambda_X^{\mu e}$ as a function of the DM mass for scalar (left panel) and tensor (right panel) bilinear structures. In the scalar case, we consider four scenarios: $X=S$, corresponding to a pure scalar interaction with $c_P^{\ell,\chi}=0$ (purple); $X=P$, corresponding to a pure pseudoscalar interaction with $c_S^{\ell,\chi}=0$ (green); and $X=S^+$ and $X=S^-$, defined by $c_S^{\ell,\chi}=\pm c_P^{\ell,\chi}$ (blue and orange, respectively). For the tensor bilinears shown in the right panel, $c_T^\chi$ is the only non-vanishing coefficient on the DM side. We present the pure tensor scenario $X=T$, corresponding to $c_{T'}^{\ell}=0$ (purple), together with the $X=T^+$ and $X=T^-$ scenarios, defined by $c_T^{\ell}=\pm c_{T'}^{\ell}$ (blue and orange, respectively). The line styles and remaining legend conventions are the same as in Fig.~\ref{fig:RMDbound}.
}
    \label{fig:boundMEGII_RMD_tensor_scalar}
\end{figure}

\medskip
Following the same procedure adopted for the vector interaction in Sec.~\ref{sec:inv_decay} and Sec.~\ref{sec:mutoegamma}, Fig.~\ref{fig:boundMEGII_RMD_tensor_scalar} shows the sensitivity to scalar (left panel) and tensor (right panel) interactions. 

\smallskip
For the scalar interactions, the purple, green, blue, and orange dashed lines represent the constraints from the invisible search for the $X=S,P,S^+,S^-$ scenarios, respectively. The blue and orange solid curves, which overlap, show the constraints obtained from the radiative analysis. In the scalar case, the invisible search provides significantly more stringent
bounds than the radiative channel. This behavior is not observed in the vector and tensor scenarios, for which the radiative process remains competitive. 

\smallskip
For the tensor interactions, the purple, blue, and orange dashed lines represent the constraints from the invisible search for the $X=T,T^+,T^-$ cases, respectively, while the blue and orange solid lines show the corresponding constraints from the $\mu\to e\chi\chi\gamma$ search. 

\medskip
To conclude this Appendix, we comment on why, for scalar interactions, the
radiative channel is significantly less sensitive than the invisible search.
The origin of this suppression can be understood through a simple
power-counting argument in the invariant mass of the invisible system.

For the invisible search, the difference between the constraints obtained
for scalar and vector interactions can be understood by considering the ratio
of the integrals in Eqs.~\eqref{eq:uscalar} and~\eqref{eq:uvector}, now
performed over $z_e\in[z_{\rm cut},1]$. This ratio provides an estimate of
the relative experimental sensitivity to scalar and vector interactions. For
the value relevant to our analysis, namely $z_{\rm cut}\sim0.76$, we find the
typical suppression
\be
\frac{\Gamma_{S}}{\Gamma_{V}}\bigg|_{z_{\rm cut}} =  \frac{1+z_{\rm cut}+z_{\rm cut}^2-3z_{\rm cut}^3}{6(1+z_{\rm cut}+z_{\rm cut}^2-z_{\rm cut}^3)} \simeq 0.09 \,,
\ee
which translates into
\be
\frac{\Lambda_{\rm scalar}}{\Lambda_{\rm vector}}
\simeq
\left(\frac{\Gamma_S}{\Gamma_V}\right)^{1/4}
\simeq 0.55\,,
\ee
in qualitative agreement with the relative strength of the invisible search limits.

For the radiative analysis, it is useful to introduce the total momentum
carried by the invisible system,
$q^\mu=(p_\chi+p_{\bar\chi})^\mu=(p_\mu-p_e-k)^\mu$.
Retaining the definition of $z_e$ and introducing the dimensionless photon
energy $z_\gamma=2E_\gamma/m_\mu$, one finds, in the muon rest frame,
\be
\frac{q^2}{m_\mu^2}
=
1-z_e-z_\gamma
+
\frac{z_e z_\gamma}{2}
\left(1-\cos\theta_{e\gamma}\right),
\ee
where $\theta_{e\gamma}$ is the physical angle between the positron and the
photon. The MEG-II radiative selection requires both the positron and the
photon to be energetic and approximately back-to-back. In this region,
$\cos\theta_{e\gamma}\simeq-1$, and therefore
\be
\frac{q^2}{m_\mu^2}
\simeq
1-z_e-z_\gamma+z_e z_\gamma
=
(1-z_e)(1-z_\gamma)\ll1\,.
\ee
Thus, the radiative selection preferentially retains endpoint-like events,
in which the visible $e^+\gamma$ system carries most of the available
energy, while the invisible pair has a small invariant mass. Averaging the
endpoint approximation above over the energy ranges defined by the cuts
$z_e\in[0.85,0.98]$ and $z_\gamma\in[0.76,1]$, one finds
$\langle q^2\rangle /m_\mu^2\simeq 10^{-2}\,.$
The radiative search therefore probes a kinematic region characterized by an
invisible invariant mass much smaller than the muon mass.

The markedly different behavior of scalar and vector interactions in this
region does not arise from the common QED bremsstrahlung structure, but is
instead determined mainly by the Dirac structure of the invisible current.
After summing over the DM spins and integrating over the two-body phase space
of the invisible pair, the scalar contribution retains an explicit factor
of $q^2$ and therefore vanishes as $q^2\to0$. For the vector interaction,
instead, the longitudinal part of the DM tensor probes the divergence of the
LFV leptonic current. Since this current connects fermions with different
masses, it is not conserved, and its divergence is proportional to
$m_\mu-m_e\simeq m_\mu$. The resulting longitudinal contribution therefore
remains finite in the small-$q^2$ limit, whereas the transverse contribution
is proportional to $q^2$ and becomes subleading. More explicitly, in the
limits $m_\chi\to0$, for equal Wilson coefficients, a common EFT scale,
and fixed visible kinematics, the scalar and longitudinal-vector
contributions satisfy
${{\rm d}\Gamma_{{\rm scalar}}^\gamma}/{{\rm d}\Gamma_{{\rm vector}}^\gamma}
\simeq 3q^2/(2m_\mu^2)$. The fiducial value
$\langle q^2\rangle/m_\mu^2\simeq10^{-2}$ therefore implies an additional
suppression of the scalar radiative signal by almost two orders of magnitude
relative to the longitudinal vector contribution. This simple power-counting
argument explains why the endpoint-oriented radiative search is substantially
less sensitive to scalar interactions than to vector ones.

One consequently expects the parametric scaling
\be
\frac{\Lambda_{\rm scalar}^\gamma}{\Lambda_{\rm vector}^\gamma}
\simeq
\left(
\frac{3\langle q^2\rangle}{2m_\mu^2}
\right)^{1/4}
\simeq
0.35 \ .
\ee
This estimate explains why the radiative bound on the scalar interaction is
weaker. By contrast, the radiative vector search remains competitive with
the corresponding invisible bound, as shown in
Fig.~\ref{fig:RMDbound}.

\section{Flavor scheme example}\label{App:FSE}
In this appendix we briefly discuss a possible flavor scheme for the various EFT operators, which is compatible with large effects in $\mu \to e \chi \overline \chi$, while providing  enough suppression for flavor-related operators, even being compatible with the very stringent constraints on LFV processes such as $\mu \to 3e$. This discussion is not meant as providing a complete model, but rather a check of plausibility.

To this extent we consider the ansatz that any charged lepton field $\ell_i$ in the effective Lagrangian is accompanied by a factor $\sqrt{y_i}$, where $y_i$ is the SM Yukawa coupling. This pattern is referred to as the Cheng-Sher Ansatz~\cite{Cheng:1987rs}, and may arise from wave-function renormalization ~\cite{Davidson:2007si, Dudas:2010yh} or partial compositeness~\cite{Keren-Zur:2012buf}, see also Ref.~\cite{Calibbi:2014yha} for a discussion in the context of supersymmetric models. In our setup we employ this pattern with the additional rule that the DM fermion $\chi$ has no suppression factor (i.e. it is flavor-blind). This scheme clearly reproducing the correct scaling of the charged lepton Yukawas, and provides the following scaling for the relevant dimension-6 operators in Eq.~\eqref{eq:FC} and \eqref{eq:lag}, giving  
\begin{align}
c_X^\ell & \sim \sqrt{y_e y_\mu} \, ,    & 
c_X^{ee} & \sim y_e \, , & 
c_X^{\mu \mu} & \sim y_\mu \, , 
\end{align}
so that for the relevant effective scales
\begin{align}
\Lambda_V^\ell & \sim 154 \Lambda \, ,    & 
\Lambda_V^{ee} & \sim 584 \Lambda  \, , & 
\Lambda_V^{\mu \mu} & \sim 41 \Lambda  \, , 
\end{align}
assuming $c_V^\chi$ is order unity. Fixing $\Lambda^\ell_V \sim 1 \TeV$ to be in the reach of MEG-II and Mu3e then requires $\Lambda \sim 6.5 \GeV$ and gives 
\begin{align}
\Lambda_V^{ee} & \sim 3.8 \TeV  \, , & 
\Lambda_V^{\mu \mu} & \sim 264 \GeV  \, , 
\end{align}
which is well within the limits from laboratory experiments, cf.~Sec.~\ref{sec:complementary}. Instead the overclosure limit on electron couplings in Eq.~\eqref{eq:analxR0ee} requires some degree of additional suppression, which corresponds to a factor about $3\div 9$ on the level of $c_X^{ee}$ depending on the reheating temperature. These estimates are somewhat in conflict with the SN limits on muons discussed in Sec.~\ref{sec:complementary}, which requires additional suppression. However, given the intrinsic uncertainty of supernova cooling limits, we consider this sector still satisfactory.    Finally, such a flavor pattern would also constrain  pure SM operators of the form 
\begin{equation}
\label{eq:SMEFT}
\mathcal{L}_{\rm LFV}
=
\frac{1}{\Lambda_{\mu e \mu e}^2}
\left(
\bar{\mu} \gamma_\mu e
\right)
\left(
\bar{\mu} \gamma_\mu e
\right)+ \frac{1}{\Lambda_{\mu e e e}^2}
\left(
\bar{\mu} \gamma_\mu e
\right)
\left(
\bar{e} \gamma_\mu e
\right)\text{h.c.}
\end{equation}
which would induce muonium-antimuonium oscillations and $\mu \to 3e$ decays. The flavor schemes gives
\begin{align}
\Lambda_{\mu e \mu e} & \sim \Lambda/\sqrt{y_e y_\mu} \sim 154 \TeV  \, , & 
\Lambda_{\mu eee} & \sim \Lambda/(y_e^3 y_\mu)^{1/4}  \sim 584 \TeV  \, , 
\end{align}
which is consistent with the limits from  muonium-antimuonium oscillations~\cite{Conlin:2020veq} and $\mu \to 3e$~\cite{Calibbi:2017uvl}
\begin{align}
\Lambda_{\mu e \mu e} & \le 5.4 \TeV  \, , & 
\Lambda_{\mu eee} & \le  207 \TeV  \, .
\end{align}
Note that dipole operators require  additional suppression, which may originate from small couplings of the heavy charged particles that induce the dipoles at loop level . Moreover, extending the flavor scheme to the $\tau$-sector requires some smallish coefficients, since one would obtain 
\begin{align}
\Lambda_V^{\tau \mu} & \sim 130 \GeV  \, , & 
\Lambda_{\tau \mu \mu \mu} & \sim 5.3 \TeV  \, , 
\end{align}
which is marginally in conflict with the $\tau \to \mu + {\rm invis.}$ partial width requiring roughly $\Lambda_V^{\tau \mu} \gtrsim 500 \GeV$ (shifting BR($\tau \to \mu + {\rm invis.}$ by 1\%) and $\tau \to 3\mu$, requiring $\Lambda_V^{\tau \mu} \gtrsim 11 \TeV$~\cite{Calibbi:2017uvl}. Finally note in the charged lepton sector a NP scale $\Lambda \sim 5 \GeV$ is sufficiently small to correspond to bosonic mediators heavy enough to provide a consistent EFT description in the charged lepton sector.

\bibliographystyle{JHEP}
\bibliography{biblio}

\end{document}